\documentclass{article}

\usepackage{microtype}
\usepackage{graphicx}
\usepackage{subfigure}
\usepackage{booktabs} % for professional tables
\usepackage[table]{xcolor}
\usepackage{multirow}
\usepackage{tablefootnote}
\usepackage{natbib}
\usepackage{times}
\usepackage{fancyhdr}
\usepackage{url}
\usepackage[colorlinks=true]{hyperref}
\usepackage{authblk} % For better author/affiliation formatting

\usepackage{amsmath}
\usepackage{amssymb}
\usepackage{mathtools}
\usepackage{amsthm}
\usepackage{dsfont}
\usepackage{algorithm}
\usepackage{algorithmic}
\usepackage{forloop}
\usepackage[capitalize,noabbrev]{cleveref}

\theoremstyle{plain}
\newtheorem{theorem}{Theorem}[section]

\theoremstyle{definition}
\newtheorem{definition}[theorem]{Definition}

\theoremstyle{remark}

\newcommand{\defeq}{\mathrel{\mathop:}=}

\newcommand{\E}{\mathbb{E}}

\newcommand{\cS}{\mathcal{S}}

\newenvironment{proof-sketch}{\noindent{\bf Proof Sketch}
  \hspace*{1em}}{\qed\bigskip\\}
\newenvironment{proof-idea}{\noindent{\bf Proof Idea}
  \hspace*{1em}}{\qed\bigskip\\}
\newenvironment{proof-of-lemma}[1][{}]{\noindent{\bf Proof of Lemma {#1}}
  \hspace*{1em}}{\qed\bigskip\\}
\newenvironment{proof-of-proposition}[1][{}]{\noindent{\bf
    Proof of Proposition {#1}}
  \hspace*{1em}}{\qed\bigskip\\}
\newenvironment{proof-of-theorem}[1][{}]{\noindent{\bf Proof of Theorem {#1}}
  \hspace*{1em}}{\qed\bigskip\\}
\newenvironment{inner-proof}{\noindent{\bf Proof}\hspace{1em}}{
  $\bigtriangledown$\medskip\\}
\newenvironment{proof-attempt}{\noindent{\bf Proof Attempt}
  \hspace*{1em}}{\qed\bigskip\\}

\usepackage[textsize=tiny]{todonotes}

\usepackage[T1]{fontenc}

\newcommand\benchname{FightLadder}

\title{\benchname{}: A Benchmark for Competitive \\Multi-Agent Reinforcement Learning}

\author[1]{Wenzhe Li}
\author[1]{Zihan Ding}
\author[1]{Seth Karten}
\author[1]{Chi Jin}
\affil[1]{Princeton University\thanks{Email: \texttt{\{wenzhe.li,zihand,sethkarten,chij\}@princeton.edu}.}}
\date{}

\begin{document}

\maketitle

\begin{abstract}
Recent advances in reinforcement learning (RL) heavily rely on a variety of well-designed benchmarks, which provide environmental platforms and consistent criteria to evaluate existing and novel algorithms. Specifically, in multi-agent RL (MARL), a plethora of benchmarks based on cooperative games have spurred the development of algorithms that improve the scalability of cooperative multi-agent systems. However, for the competitive setting, a lightweight and open-sourced benchmark with challenging gaming dynamics and visual inputs has not yet been established. In this work, we present \benchname{}, a real-time fighting game platform, to empower competitive MARL research.
Along with the platform, we provide implementations of state-of-the-art MARL algorithms for competitive games, as well as a set of evaluation metrics to characterize the performance and exploitability of agents. We demonstrate the feasibility of this platform by training a general agent that consistently defeats 12 built-in characters in single-player mode, and expose the difficulty of training a non-exploitable agent without human knowledge and demonstrations in two-player mode. 
\benchname{} provides meticulously designed environments to address critical challenges in competitive MARL research, aiming to catalyze a new era of discovery and advancement in the field.
Videos and code at \href{https://sites.google.com/view/fightladder/home}{https://sites.google.com/view/fightladder/home}.
\end{abstract}

\section{Introduction}

As an active branch of artificial intelligence (AI), deep reinforcement learning (DRL) has achieved significant success in various domains, including, but not limited to, strategic games~\citep{silver2016mastering,li2020suphx,moravvcik2017deepstack,vinyals2019grandmaster,berner2019dota}, robotics control~\citep{lillicrap2015continuous,andrychowicz2020learning,brohan2022rt}, and large language models alignment~\citep{ouyang2022training}.
Underpinning these rapid advances are not only the development of sample-efficient RL algorithms but also the availability of well-designed benchmarks. These benchmarks provide environmental platforms, unify evaluation protocols, enable comparisons of state-of-the-art methods, motivate improved solutions, and guide practical applications.
As an example, policy proximal optimization (PPO)~\citep{schulman2017proximal} demonstrates its superior performance across different single-agent RL benchmarks, hence being considered as one of the most widely adopted single-agent RL algorithms~\citep{andrychowicz2020matters}. 
In the realm of multi-agent reinforcement learning (MARL), while a series of benchmarks have also been proposed, most of them focus on fully cooperative settings. For competitive environments, some platforms simulate games with tabular representations and relatively simple dynamics, such as board games, while others, based on complex game engines, require significant computational resources and expert knowledge, such as Starcraft II and DOTA. To advance research on competitive multi-agent reinforcement learning (MARL) and transform game-theoretical results into practical applications, a fully competitive game platform that strikes the right balance between complexity, efficiency, and generality is urgently needed.

\begin{figure}[!t]
% \vskip 0.2in
\begin{center}
\centerline{\includegraphics[width=\textwidth]{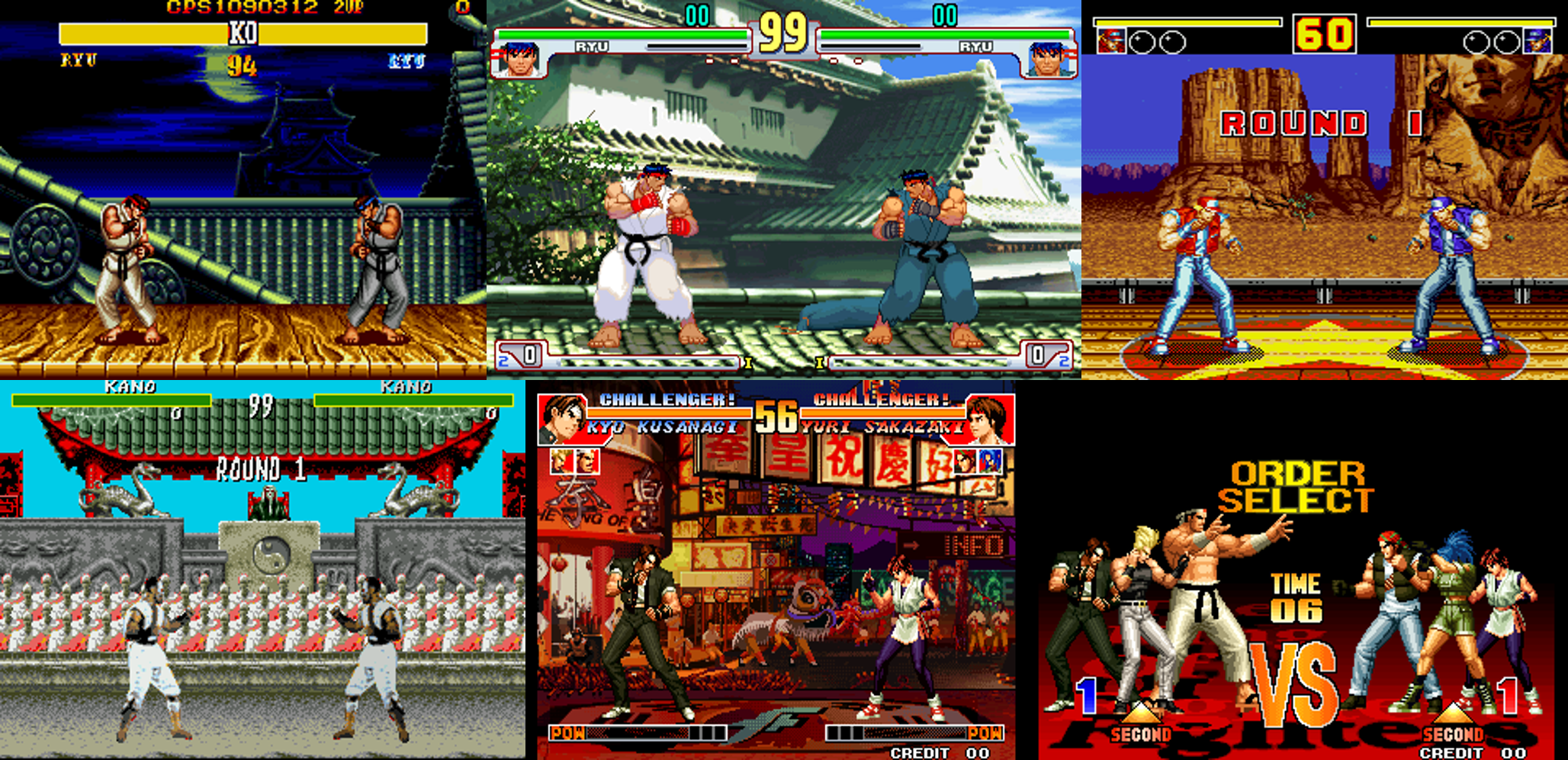}}
\vskip -0.1in
\caption{\benchname{} currently supports various cross-platform video fighting games: \textit{Street Fighter II} (Genesis platform), \textit{Street Fighter III} (Arcade platform), \textit{Fatal Fury 2} (Genesis platform), \textit{Mortal Kombat} (Genesis platform), and \textit{The King of Fighters '97} (Neo Geo platform).}
\label{fig:scenarios}
\end{center}
\vskip -0.4in
\end{figure}

Multi-agent games are known to be more challenging than single-agent ones due to the additional non-stationarity introduced by the interactions with other players. Among different types of interactions, fully competitive settings can be rather difficult. People have a long history of designing and playing competitive games, as well as building strong AI opponents
to make the game more challenging and hence intriguing. Previous AI research has investigated the solutions of competitive games using RL, but mostly for small-scale games like Backgammon~\citep{tesauro1995temporal} or other board games~\citep{schrittwieser2020mastering, brown2018superhuman, brown2019superhuman}. Moreover, this line of work mostly uses state vectors as inputs, which is arguably easier than directly learning from raw pixel inputs that commonly appear in most popular video games. 
In contrast, this paper considers fighting games, which feature rich policy space, and significant depth in strategy --- including catching specific timing, counter-attack by exploiting the stiffness of the opponents, managing energy resources, etc. Moreover, these games also have a rather large number of characters with distinct move-sets which add another layer of complexity for AI agents to master the game.
As a result, we are motivated to build a platform for a series of fighting games, with image inputs and complex fighting dynamics, to serve as a challenging competitive multi-player platform for the broad AI research community.

Apart from the game platform, the evaluation criteria and benchmark results for certain game settings are essential for boosting the field. MARL has been greatly investigated in the past few years for solving multi-player games, from both theoretical and empirical perspectives. A large number of algorithms have been proposed according to specific settings~\citep{sunehag2017value, yu2022surprising, lowe2017multi, silver2018general, lanctot2017unified, vinyals2019grandmaster, ding2022deep}. Nonetheless, for competitive game settings, there is a lack of unified evaluation criteria with thorough comparisons among different approaches.

In this work, we present \benchname{}, a competitive two-player games benchmark. Our contributions are three-fold: We build the \benchname{} platform to support five two-player fighting games, with ease to extend to other games in the future. The games support various observation spaces involving rendered images. Based on prior work, we provide implementations of the most popular algorithms for solving these competitive games, including an AlphaStar league training algorithm~\citep{vinyals2019grandmaster} and policy space response oracle~\citep{lanctot2017unified}. Furthermore, a unified evaluation framework with \textit{Elo rating} and \textit{exploitability} tests are provided alongside the game platforms and algorithm library. We report experimental results using the above toolkits to serve as the baselines for two-player competitive game settings. 
One important challenge of MARL is its diverse nature, which includes collaborative games, competitive games, two-player games, and multiplayer games, all of which have rather different problem structures, properties, and solution concepts. While it is promising to develop a unified solution that addresses them all together, in this work, we empirically demonstrate that  \textit{to some extent, existing methods are still limited in solving competitive two-player zero-sum games alone when combined with visual input, rich strategy space, and lack of extensive human demonstration}. We hope that \benchname{}, which particularly focuses on this fundamental two-player setting, can serve as a stepping stone for the research community to develop effective self-play style algorithms to tackle it first before moving on to even more complicated scenarios, and inspire future directions that involve more types of interactions.

\section{Related Work}

\paragraph{MARL Environments.}
MARL environments can be categorized into three types according to the payoff structure of the game: \textit{fully cooperative}, \textit{fully competitive}, and \textit{general}.

Existing environments for \textit{fully cooperative} games are designed for various scenarios, including simulated games like MAMuJoCo~\citep{peng2021facmac}, card games like Hanabi~\citep{bard2020hanabi}, video games like small-scale StarCraft SMAC~\citep{samvelyan2019starcraft} and Google Research Football~\citep{kurach2020google}, as well as practical scenarios like Traffic Junction~\citep{sukhbaatar2016learning} in a grid world, Flatland~\citep{mohanty2020flatland} for railway networks, network load balancing~\citep{yao2022learning} and CityFlow~\citep{zhang2019cityflow} for city traffic. Cooperative environments feature a single reward function shared by all agents, which makes them distinct from competitive games.

On the other hand, the \textit{fully competitive} game benchmarks are relatively underdeveloped. Prior competitive environments are either on games with low-dimensional or discrete state space such as Pommerman~\citep{resnick2018pommerman} and board games \citep{tesauro1995temporal,schrittwieser2020mastering, brown2018superhuman}; or complex games with image input that require a significant amount of computational resources, such as Starcraft II~\citep{vinyals2019grandmaster} or DOTA~\citep{berner2019dota}. The fighting game environments proposed in this paper strike the right balance between complexity, efficiency, and generality. A few previous works also have explored fighting games: \citet{go2023phase} focuses on developing an algorithm for a single fighting game---street fighter, as opposed to this paper which provides an environment that supports various fighting games. While \citet{palmas2022diambra} provides a platform for fighting games, most of its efforts have been focused on the single-agent setting. It lacks explicit criteria for two-player scenarios with adaptive opponents, and does not provide a benchmark evaluating existing competitive MARL algorithms. \citet{khan2022darefightingice} focuses on fighting games in the blind setting where agents have to rely on acoustic inputs to play.

Finally, there are also a number of environments for \textit{general} multiagent games that feature both cooperation and competition, including MPE~\citep{mordatch2018emergence}, MAgent~\citep{zheng2018magent}, Hide-and-Seek~\citep{baker2019emergent}, DMLab2D~\citep{beattie2020deepmind}, Arena~\citep{song2020arena}, Smarts~\citep{zhou2020smarts}, Neural MMO~\citep{suarez2021neural}, PettingZoo~\citep{terry2021pettingzoo}, MATE~\citep{pan2022mate}, etc. Generic multi-agent general-sum games are rather challenging to evaluate --- even the optimal solution concepts remain elusive. In contrast, the fully competitive setting considered in this paper presents clear game-theoretic properties and well-defined solution concepts. We also remark that while a number of the platforms above support several fully competitive games, they did not provide carefully designed evaluation toolkits as well as extensive baselines for competitive MARL algorithms.

% \vspace{-0.2in}
\paragraph{MARL Algorithms and Evaluation.}
% Algorithms and Libraries
To solve multi-agent learning tasks, researchers have proposed algorithms and built libraries for ease of usage and evaluation. PyMARL~\citep{samvelyan2019starcraft} is an initial MARL library built for solving SMAC tasks, while PyMARL2~\citep{hu2021rethinking} extends PyMARL with QMIX~\citep{rashid2020monotonic}. EPyMARL~\citep{papoudakis2020benchmarking} is also an extension of PyMARL, as a unified library for cooperative games supporting different learning paradigms including centralized and decentralized learning, value decomposition, etc. MARLlib~\citep{hu2023marllib} includes major cooperative MARL algorithms like VDN~\citep{sunehag2017value}, MAPPO~\citep{yu2022surprising}, MADDPG~\citep{lowe2017multi}, etc. More recent libraries include Pantheonrl~\citep{sarkar2022pantheonrl}, MAlib~\citep{zhou2023malib}, etc. These libraries mainly support MARL algorithms for cooperative games, lacking support for solving competitive games. 

% Self-Play, Population-Based Methods, and League Training.
On the other hand, there is a line of research for solving competitive games with algorithms like self-play~\citep{silver2018general}, fictitious play~\citep{brown1951iterative}, Nash Q-learning~\citep{hu2003nash, ding2022deep}, double oracle~\citep{mcmahan2003planning}, policy space response oracle (PSRO)~\citep{lanctot2017unified} and league training~\citep{vinyals2019grandmaster}. A unified benchmark remains missing to compare and evaluate the efficiency these algorithms on the same set of tasks, especially when combined with deep RL. This paper addresses this issue in the fully competitive setting. We concentrate on two-player zero-sum games, and propose a platform for fighting-style fully competitive games, along with a baseline implementation and evaluation of popular algorithms.

\section{Multi-Agent Reinforcement Learning}
\label{sec:marl}
\benchname{} is designed to motivate novel algorithms for fully competitive two-player games in the domains of MARL and game theory. 
Markov Games (MGs) \citep{shapley1953stochastic} generalize single-player Markov Decision Processes (MDPs) into multi-player settings. Each player has its own utility and optimizes its policy to maximize the utility. The two-player zero-sum setting in MG represents a competitive relationship between the two players. With a shaped dense reward, the games can be generalized to general-sum. 

We denote a finite-horizon two-player general-sum partially observable MG as $\mathrm{POMG}(\mathcal{S}, \mathcal{O}, \mathcal{A}, \mathcal{B}, \mathbb{P}, \mathbb{O}, \{r\}_{i=1}^2, H)$. $\mathcal{S}$ is the state space, which can be partially observable and transformed through an observation emission function $\mathbb{O}$: $\mathcal{S}\rightarrow \mathcal{O}$ to the observation space $\mathcal{O}$. $\mathcal{A}$ and $\mathcal{B} $ are action spaces for two players, respectively. $\mathbb{P}( \cdot | s, a, b)$ is the state transition distribution, $r_i: \mathcal{S} \times \mathcal{A} \times \mathcal{B} \rightarrow \mathbb{R}$ is the reward function for the $i$-th player. In the zero-sum setting, two reward functions satisfy the zero-sum payoff structure $r_1+r_2=0$. $H$ is the horizon length. We denote the policies of two players as $\mu$ and $\nu$, respectively. $V_i^{\mu, \nu} \colon \cS \to \mathbb{R}$ represents the value function for player $i$ evaluated with policies $\mu$ and $\nu$, which can be expanded as the expected cumulative reward starting from the state $s$, 
\begin{equation*} 
\textstyle	 V_i^{\mu, \nu}(s) 
\defeq \E_{\mu, \nu}\big[\sum_{h =
        1}^\infty r_i(s_{h}, a_{h}, b_{h}) \big| s_1 = s\big].
\end{equation*}
In zero-sum games, we have $V_1^{\mu, \nu}(s)=-V_2^{\mu, \nu}(s), \forall s\in \mathcal{S}$ and define $V^{\mu, \nu}(s)=V_1^{\mu, \nu}(s)$ for simplicity.

\begin{definition}[Best Response]
For any policy of the first player $\mu$, there exists a \emph{best response} (BR) against it from the second player, which is a policy
$\nu^\dagger(\mu)$ satisfying $V_{2,h}^{\mu, \nu^\dagger(\mu)}(s) = \max_{\nu} V_{2, h}^{\mu, \nu}(s)$ for
any $(s, h) \in \cS \times [H]$. 
We denote $V_{2,h}^{\mu, \dagger} \defeq V_{2,h}^{\mu, \nu^\dagger(\mu)}$ for simplification. $V_{2,h}^{\mu, \nu}(s)$ is the value function of the second player. BR against the second player can be defined similarly.
\end{definition}

\begin{definition}[Nash Equilibrium]
    The \emph{Nash equilibrium} (NE) in zero-sum setting is defined as a pair of policies $(\mu^\star,\nu^\star)$ 
satisfying the following minimax equation:
\begin{equation*}
\max_{\mu} \min_{\nu} V^{\mu, \nu}(s) = V^{\mu^\star, \nu^\star}(s) = \min_{\nu} \max_{\mu} V^{\mu, \nu}(s).
\end{equation*}
\end{definition}

\begin{definition}[Exploitability]
\label{def:exploitability}
The exploitability for a policy $\mu$ of the first player is defined as 
$V_2^{\mu, \dagger}(s_1) - V_2^{\mu^\star, \nu^\star}(s_1) $, i.e., the value of its BR policy $\nu^\dagger(\mu)$ or the suboptimality gap from the NE value.
The exploitability of the other side policy $\nu$ can be defined accordingly.
\end{definition}

Note that NE strategies will always lead to zero exploitability, thus approaching the non-exploitable strategies is a reasonable pursuit for the game.

\section{\benchname{}} \label{sec:scenarios}
In this section, we present technical details of \benchname{}. In the following part, we first introduce different game settings of \benchname{}, followed by elaborating elements of MGs corresponding to the environment, and conclude with highlighting features of our benchmark.

\subsection{Scenarios}

\benchname{} provides a flexible interface between modern game emulators~\citep{murphy2013hacking, nichol2018retro} and algorithm developers. Thanks to its flexibility, \benchname{} can support a wide range of classical fighting games over the past decades, including Street Fighter, Mortal Kombat, Fatal Fury, and The King of Fighters, some of which are still very popular nowadays. Figure~\ref{fig:scenarios} shows screenshots of several fighting games provided by \benchname{}. With this diverse set of supported games, we can benchmark algorithms on various fighting scenarios differing in backgrounds, characters, and moving dynamics, which can further motivate novel algorithms that are general rather than overfitting to one specific game. For better readability and clarity, we would use Street Fighter as an example for illustration and evaluation in the rest of the paper. The other fighting games are very similar, and readers could refer to Appendix~\ref{appx:scenarios} for more details. We name each scenario in the form \emph{[game alias]\_[character left]\_vs\_[character right]}, for example \emph{sf\_ryu\_vs\_ryu} in Street Fighter.

While \benchname{} mainly focuses on the competitive two-player setting, the nature of fighting games allows it to be seamlessly deployed to the single-player scenario where the agent's task is to compete against a built-in game AI (e.g., \emph{sf\_ryu\_vs\_ryu(cpu)}). Under this single-player setting, users have the freedom to choose characters and set up the difficulty of the scripted AI opponent. Moreover, our benchmark also supports training in a much more challenging full-game scenario (e.g., \emph{sf\_ryu\_full\_game}), where the agent needs to defeat all 12 characters controlled by computers with the difficulty progressively increasing. As we shall see in later experiments, this scenario could also serve as a sanity check for our baseline algorithms to see whether they could learn effective behaviors from the environment.

\subsection{State and Observations}
We define the state space $\mathcal{S}$ as the complete set of attributes stored in the game emulator after each step of action. Same as human players, the agent is not allowed to access the underlying full state but can only access the observation space $\mathcal{O}$ of pixels, which forms a 128$\times$100 RGB image corresponding to the rendered screen. This image includes the position and movement of both sides of the players, as well as the hit-point bar and the round timer on the top of the screen. At every step, a configurable number of images are stacked as the input of the agent. 

While we use pixels as default observations, we also provide an interface for users to access additional information about the game status, including position, hit-point, and exact countdown number for agents on both sides. Users can leverage these attributes to better understand the agent's behavior or augment feature representations. More details are provided in Appendix~\ref{appx:scenarios}.

\subsection{Action Space}
\begin{figure}[t]
% \vskip 0.2in
% \begin{center}
\hskip -0.1in
\includegraphics[width=0.5\columnwidth]{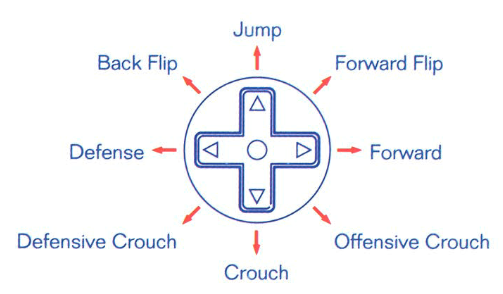}
\includegraphics[width=0.5\columnwidth]{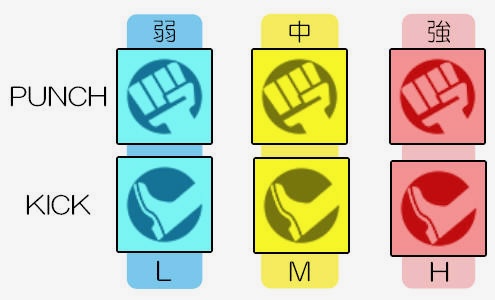}
\vskip -0.1in
\caption{Motion and attack action spaces of fighting games. Images are adapted from Instruction Manual of Street Fighter II.}
\label{fig:actions}
% \end{center}
\end{figure}

In fighting games, two players share the same action space $\mathcal{A}$.
The native \emph{human action space} $\mathcal{A}_\text{human}$ is designed to mimic the joystick control of arcade games, which is a 12-dimensional binary space (['B', 'A', 'MODE', 'START', 'UP', 'DOWN', 'LEFT', 'RIGHT', 'C', 'Y', 'X', 'Z']) with each dimension representing a button being pressed or not. Note that due to the nature of fighting game engines, this space contains many redundant actions that are invalid, for instance, moving in opposite directions or moving and attacking at the same moment. To filter out these redundant actions and to construct a more structured space, we develop a categorical \emph{transformed action space} 
$\mathcal{A}_\text{trans}$ through an encoding function $F:\mathcal{A}_\text{human} \rightarrow \mathcal{A}_\text{trans} $. Specifically, $\mathcal{A}_\text{trans}$ is the joint set of a direction move set $\mathcal{A}_\text{motion}$=\{defense, forward, jump, crouch, back flip, front flip, offensive crouch, defensive crouch\} and an attack move set $\mathcal{A}_\text{attack}$=\{light punch, medium punch, hard punch, light kick, medium kick, hard kick\}, as shown in Figure~\ref{fig:actions}. Each action will remain a number of frames according to users' configuration. The games also have special techniques called \textit{close attack}, i.e., Throws and Holds, which can be applied in certain regions near the opponent.

\begin{figure}[t]
\vskip -0.1in
\begin{center}
\includegraphics[width=0.3\columnwidth]{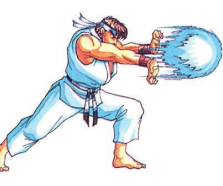}
\includegraphics[width=0.3\columnwidth]{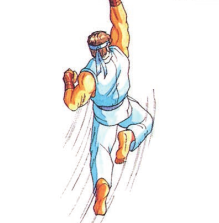}
\includegraphics[width=0.3\columnwidth]{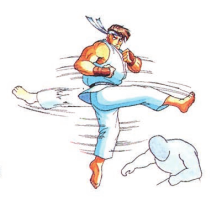}
\vskip -0.1in
\caption{Example of special moves for character Ryu in StreetFighter II (left to right): Fireball, Dragon Punch, Hurricane Kick. Images are adapted from Instruction Manual of Street Fighter II.}
\label{fig:special_move}
\end{center}
\vskip -0.3in
\end{figure}

In addition to the standard move set, one signature element of fighting games is \textit{special moves}, which is a kind of powerful attack or maneuver that requires the player to follow a specific action sequence (i.e., sequential keys combination, or combination of key holding and key pressing), with an example depicted in Figure~\ref{fig:special_move}. These moves usually have special properties (e.g., invincibility frames, larger coverage, etc.) and play a critical role in the strategy and depth of the game. They are especially useful for higher levels of play, from which players could create complex combos and outperform opponents. However, we observe that learning to perform special moves from scratch can be challenging to baseline algorithms, as it requires the agent to memorize frames and actions in previous steps and accurately perform the next action in the action sequence of special moves. Moreover, the special moves can be different from character to character, which increases the difficulty of the game. Therefore, to alleviate this challenge, we also include hard-coded special move lists as one part of the action space so that the agent can directly access special moves with one single action. 

\subsection{Rewards}
\paragraph{Sparse Reward.}
Both sides of the agents are to maximize their win rate for each round of the game. The \emph{sparse reward} $r_\text{sparse}$ assigns +1 for the winner and -1 for the loser at the end of each episode. In the sparse reward setting, all fighting games are two-player zero-sum
games, which are theoretically guaranteed to exist at least one Nash Equilibrium~\citep{filar2012competitive}, which directly induces a pair of non-exploitable policies.

\paragraph{Win Rate.}
For two players $A$ and $B$, policy $\pi_A$ winning against policy $\pi_B$ can be defined as a reward relationship $r^{A}_\text{sparse}(\pi_A, \pi_B) > r^{B}_\text{sparse}(\pi_A, \pi_B)$ in a single match, with $r^{A}_\text{sparse}$ and $r^{B}_\text{sparse}$ as the sparse reward for players A and B in the zero-sum setting. The win rate is defined as the probability of winning as $p(\pi_A\succ \pi_B)$.

\paragraph{Shaped Dense Reward.} While sparse reward is straightforward for evaluation, we discover that baseline algorithms could not effectively learn to behave well from such a sparse signal. To address this issue, we introduce a \emph{shaped dense reward} $r_\text{dense}$ for training, which is a weighted sum of the hit-point damage inflicted by the agent on the opponent and the damage it receives, together with a bonus (penalty) for winning (losing) the game. Specific format of this reward refers to Appendix~\ref{appx:dense_reward}. The dense reward $r_\text{dense}$ is chosen to coincide with the win rate of the policy, such that $\pi_A\succ \pi_B$ will always lead to $ r^{A}_\text{dense}(\pi_A, \pi_B) > r^{B}_\text{dense}(\pi_A, \pi_B)$ in expectation. The dense reward also offers some flexibility,
that the user can control the agent's aggressiveness by configuring the weighing scales in the reward function.

\subsection{Features}
We remark on the following features of the proposed benchmark that could benefit MARL research.

\paragraph{Rich Strategy Space.} 
One key feature of our benchmark is the rich strategy space as the nature of fighting games, which is particularly beneficial to the development of game-theoretical algorithms. To name a few, fighting games require players to consider \textbf{(a) character diversity:} each character has a unique skill set with different strengths and weaknesses, so one needs to master the strategy and counter-strategy of all possible opponents, and even reason how to select and order characters when they have the freedom to do so; \textbf{(b) complexity of mechanics:} fighting games are designed with sophisticate mechanics such as invincibility frame, hitboxes, and combo systems, which are challenging for micromanagement of characters; and \textbf{(c) adversarial opponents:} opponents may progressively adapt their policies to players' policies, thus finding non-exploitable policies is crucial in mastering fighting games. 

\paragraph{Various Difficulty Levels.}
FightLadder provides several kinds of scenarios: single-player mode against one CPU player (e.g., \emph{sf\_ryu\_vs\_ryu(cpu)}), single-player mode full game (e.g., \emph{sf\_ryu\_full\_game}), two-player mode (e.g., \emph{sf\_ryu\_vs\_ryu}), team mode (supported in some games such as The King of Fighters). The difficulty levels are increasing in this order, as two-player mode (no CPU) introduces additional non-stationary (opponents can be adaptive), and team mode offers a richer strategy space. Moreover, FightLadder supports specifying \textbf{arbitrary difficulty levels} of CPUs and \textbf{arbitrary characters} for both the player and its opponent. This enriches the features of our platform and the diversity of strategy space.

\paragraph{Computational Efficiency.}
\benchname{} also enjoys efficient computation for its usage, and the comparison with several other popular game environments is shown in Table~\ref{tab:fps}. The frame rate is 13 times faster than SMACv2, with one-fourth usage of the memory. While \benchname{} is less efficient than PettingZoo Atari, it provides more game complexity. The balance of complexity and low computational cost is important for evaluating algorithms at scale. 

\begin{table}[t]
\vskip -0.1in
  \centering
  \caption{FPS and memory usage of several open-sourced platforms.}
  \vskip 0.1in
  \label{tab:fps}
  % \resizebox{\columnwidth}{!}{
  \begin{tabular}{lccc}
    \toprule
    \textbf{Environment} & \textbf{Speed (FPS)} & \textbf{Memory (MB)} \\
    \midrule
    \benchname{} (Ours) & 1935.76 & 195.46 \\
    SMACv2 & 146.72 & 876.96 \\
    PettingZoo Atari & 6268.18 & 32.13\\
    DMLab2D & 1144.27 & 47.41 \\
    \bottomrule
  \end{tabular}
  % }
\vskip -0.1in
\end{table}

\paragraph{Fidelity and Popularity.}
\benchname{} allows testing agents in full-length fighting games with an interface similar to human perception, thus providing a high-fidelity evaluation of competitive RL algorithms. Moreover, fighting games have been gaining popularity since they were released, making it easier to test the learned RL agents against human expert players. 

\paragraph{Open-Source and Compatibility.}
\benchname{} is designed for the broad RL research community, so we make efforts to improve the ease of usage and make it accessible to all potential users. It is compatible with the Gym~\citep{brockman2016openai} interface so that users can leverage off-the-shelf RL algorithms implementation. 

\paragraph{Customization, Extension, and Flexibility.}
\benchname{} is extremely flexible for configuration and extension. For customization, the users can customize action spaces (human/transformed action), reward functions (sparse/tunable shaped dense reward), number of frames to be observed per step, as well as access to additional information to help training. Moreover, our platform is built upon popular modern game emulators so that it is easy to extend to other games not provided by us. Specifically, it supports \href{https://github.com/openai/retro}{Gym Retro} and \href{https://github.com/M-J-Murray/MAMEToolkit}{MAMEToolkit}, which already support a wide range of games. This extension capability of
diverse games is provided by our platform with minimal engineering efforts. Please check our \href{https://sites.google.com/view/fightladder/home}{open-source project}\footnote{\href{https://sites.google.com/view/fightladder/home}{https://sites.google.com/view/fightladder/home}} for more details.

\section{Evaluation Metrics} \label{sec:metrics}
\paragraph{Versus Built-In Game AIs.} 
Directly competing with the built-in AIs of the games provides a straightforward way of measuring policy performance.
Typically, fighting games offer a hierarchical structure of levels, enabling players to adjust the difficulty setting (for example, Street Fighter features eight distinct levels). This structure allows for the empirical evaluation of the policy against the game's scripted AI at varying levels of challenge. It is important to acknowledge, however, that the limitations associated with hard-coded adversaries restrict the extent to which this metric can accurately reflect the policy's real capability. For brevity, we shall refer to such agents as CPU.

\paragraph{Elo Ratings.} 
The skills of agents can be ranked through the FIDE rating system~\citep{elo1978rating}, which is an incremental learning system that increases the Elo of winners and decreases the Elo of losers.
The larger the difference in Elo between players $A$ and $B$, the higher the probability that the player with the higher Elo, $A$, beats the player with the lower Elo, $B$. The Elo score calculation takes the following procedures:

First, the probability of player $A$ winning is estimated with,
\begin{equation*}
   p_A:= p(\pi_A \succ \pi_B) = ( 1.0 + 10^{\frac{\text{Elo}_B - \text{Elo}_A}{400}} )^{-1}.
\end{equation*}
Then the Elo rating for player $A$ as $\text{Elo}_A$ will be updated with following formula:
\begin{equation*}
    \text{Elo}_A = \text{Elo}_A + k\cdot (\mathds{1} [\text{winner} = A] - p_A),
\end{equation*}
where $k$ is a constant of update rate.
The update is symmetric for player $B$, as well as any other player in the ranking system.

\paragraph{Versus AI Exploiters.} 
As discussed in Section~\ref{sec:marl}, exploitability (as Definition~\ref{def:exploitability}) measures the distance of a policy to the Nash equilibrium of the game. Specifically, the exploitability of a policy $\mu$ is measured by the win rate of its BR policy $\nu^\dagger(\mu)$ against $\mu$, since $V^{\mu^\star, \nu^\star}(s_1)=0$ for symmetric zero-sum game and  $V_2^{\mu, \dagger}(s_1)= 1 \cdot p(\nu \succ\mu) + 0 \cdot p(\nu \preceq \mu) = p(\nu \succ\mu)$ for sparse reward setting.
In practice, we can use any single-agent deep RL algorithm as an exploiter to approximately learn the BR policy $\nu^\dagger(\mu)$. For fair comparisons, we should use one consistent exploiter (same RL algorithm with same configurations) to evaluate the exploitability of different baselines.

\paragraph{Versus Human Players.} 
While Definition~\ref{def:exploitability} is a general metric to measure exploitability, it may be limited to the capability of deep RL algorithms in usage. Therefore, we also provide an interface for human players such that they can play with any learned model with convenience. This feature will show the strengths and weaknesses of agents directly and visibly, and motivate developers to improve their algorithms to be more non-exploitable in general. Given the remarkable success of modern RL algorithms outperforming expert human players in various video games~\citep{mnih2013playing, vinyals2019grandmaster, berner2019dota}, we believe that \benchname{} will emerge as a promising platform for the broad competitive MARL community and researchers will eventually build AI agents that could beat world champions in a much richer set of strategic games with significantly less engineering efforts.

\section{\benchname{}-Baselines} \label{sec:baselines}
For the convenience of the community to evaluate existing methods and new algorithms on FightLadder platform, we open-source the implementation of several state-of-the-art (SOTA) competitive MARL algorithms,
including independent learning~\citep{de2020independent}, two-timescale learning~\citep{daskalakis2020independent}, fictitious self-play~\citep{pmlr-v37-heinrich15}, policy-space response oracle~\citep{lanctot2017unified} and league training~\citep{vinyals2019grandmaster}. Our codebase supports decentralized learning across multiple GPUs, and it is built upon Stable-Baselines3~\citep{stable-baselines3} so that users can leverage off-the-shelf implementations of RL algorithms. We choose proximal policy optimization (PPO)~\citep{schulman2017proximal} as the backbone policy optimization algorithm in our experiments. More details of baseline algorithms refer to Appendix~\ref{appx:baselines}.

\begin{figure*}[htbp]
% \vskip 0.2in
\begin{center}
\centerline{\includegraphics[width=\textwidth]{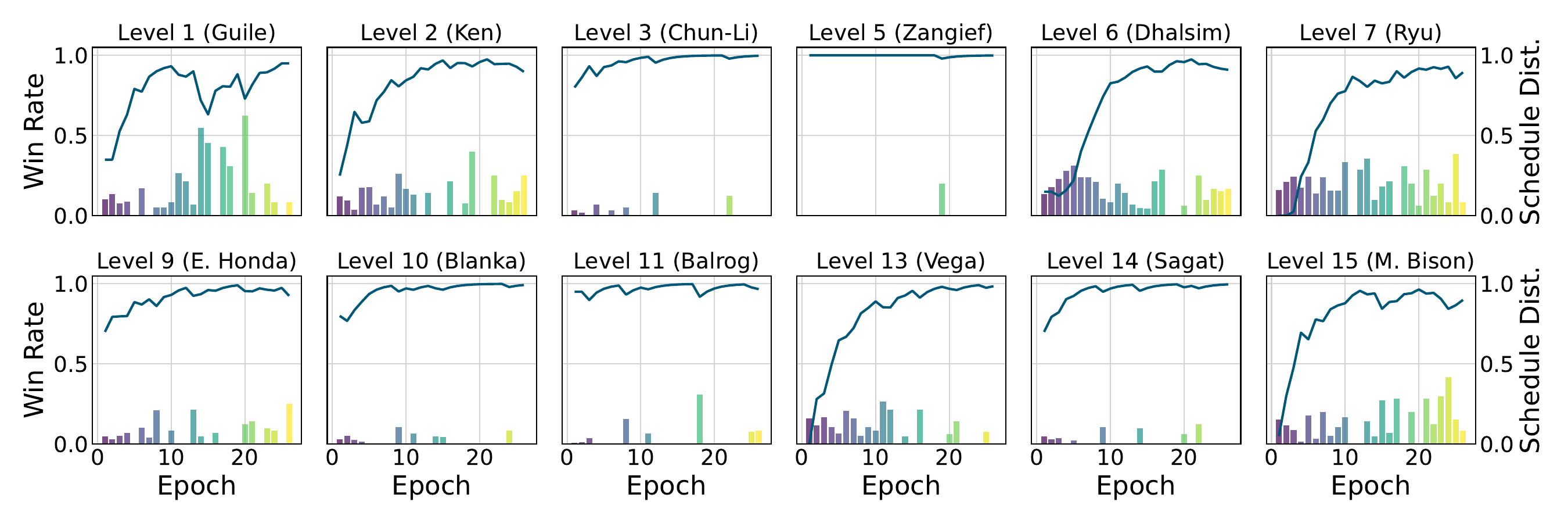}}
% \vskip -0.2in
\caption{The win rate curves and scheduling distribution bar plot in \emph{sf\_ryu\_full\_game} via the proposed PPO with curriculum learning. Opponents of different characters are marked with different levels. Levels 4, 8, and 12 are omitted as they are bonus levels without fighting.}
\label{fig:1p_win_rate}
\end{center}
% \vskip -0.3in
\end{figure*}

\section{Results}
In this section, we provide benchmark results on a selected game in \benchname{}--the Street Fighter. We aim to answer the following questions through our benchmark: \textbf{(a)} Can existing RL algorithms solve the full video game in the single-player scenario? \textbf{(b)} How does the performance of state-of-the-art baseline algorithms in the two-player competitive setting compare? and \textbf{(c)} Does multi-agent training help to improve the non-exploitability? 

\subsection{Single-Player Full Video Game}
To answer question \textbf{(a)}, we evaluate PPO's performance in the scenario \emph{sf\_ryu\_full\_game} with human action space as a feasibility check. As mentioned in Section~\ref{sec:scenarios}, this scenario requires the agent to learn a generalizable policy to compete against all different characters with increasing difficulty levels. \textit{Curriculum learning} is applied to train the policy from easy to hard cases. Furthermore, to improve learning efficiency we develop a curriculum scheduler for opponent sampling to match with the learner after each epoch. More specifically, for the current learner $L$ with policy $\pi_L$, we sample its opponent $C$ from the entire character set $\mathcal{C}$, with the following inverse-weight scheduling distribution: $$C\sim \Delta(\mathcal{C})\propto 1-p(\pi_L\succ \pi_C),$$ where $p(\pi_L\succ \pi_C)$ is the win rate of the learner against the opponent and $\Delta(\cdot)$ is the simplex. Intuitively, such a curriculum will encourage the agent to focus on the hardest opponents, similarly to prioritized experience play~\citep{schaul2015prioritized}. We defer other implementation details to Appendix~\ref{appx:implementation}.

Figure~\ref{fig:1p_win_rate} shows the performance of our proposed method during training. With 20 epochs of training (each epoch involves 10M training steps competing with opponents sampled from the curriculum scheduler in parallel), the agent is capable of defeating characters at each level with a win rate close to 1. In addition to beating each character with a high probability, the trained policy can complete the full video game with over 0.6 win rate, outperforming human players with hours of playing experience. This result shows that existing RL algorithms can already learn a well-behaved policy to solve the full single-player video game, which provides a good starting point for exploring the multi-agent setting.

As an additional experiment, we also test the inclusion of hard-coded special move lists in this setting with exactly the same algorithm implementation. Although it could be easier for the agent to learn more offensive policies, significant improvement in the overall win rate is not observed. It indicates that the agents without encoded special moves can also effectively learn policies against CPUs. Constantly playing special moves will lead to a vulnerable situation for the agent, whereas the defensive strategy also matters greatly in the game. Moreover, given that an experienced human player can perform special moves easily (by executing the action sequences almost instantly), we do not think that hard-coded special move lists will become the advantage of trained agents over human players.

\subsection{Performance of Two-Player Baseline Algorithms} \label{sec:2p_exp}
\begin{figure}[!h]
    \centering
    \includegraphics[trim=5cm 0.5cm 5cm 0.5cm,clip,width=\columnwidth]{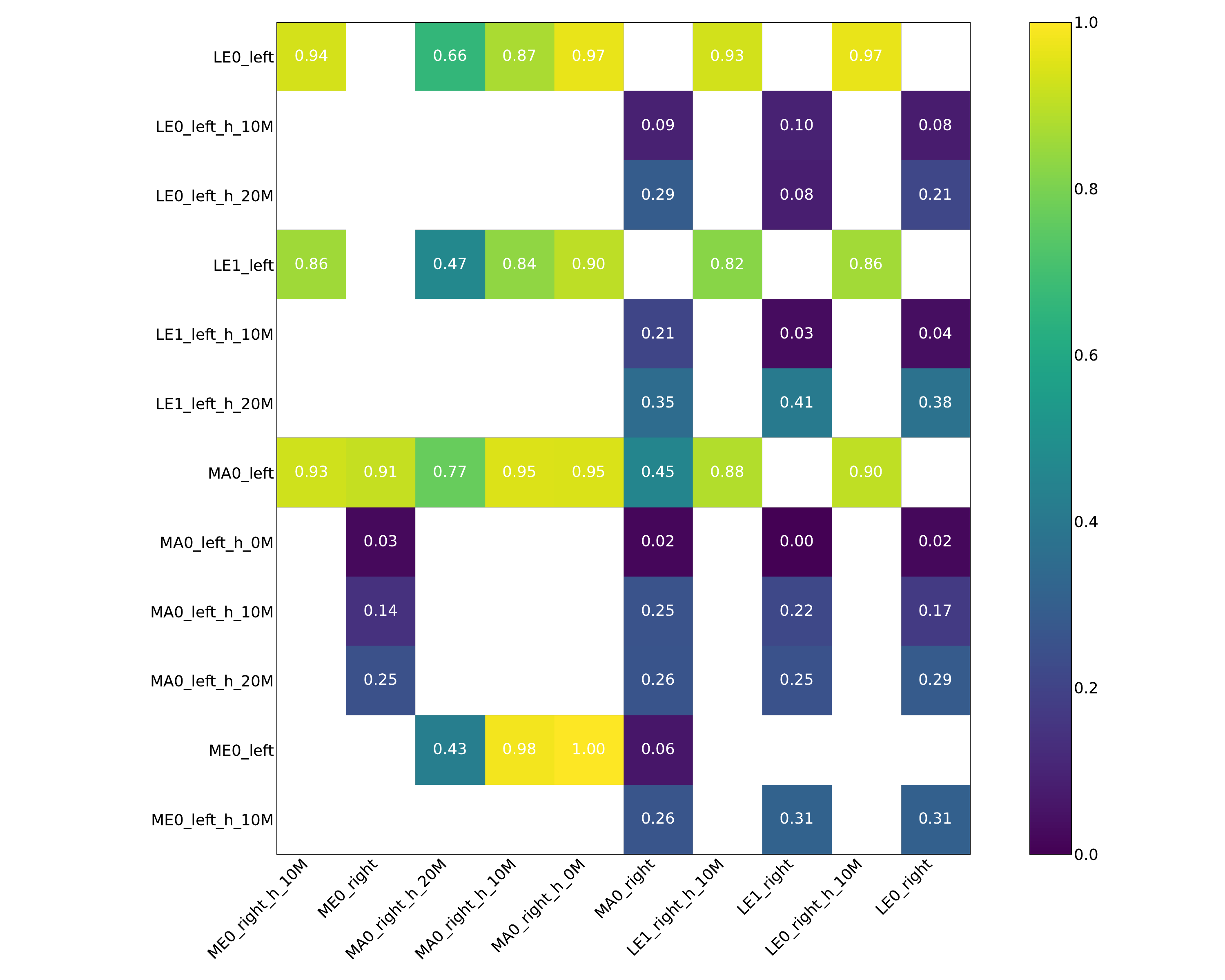}
    \caption{The payoff matrix for each pair of agents at a certain stage of League training. For league training, there is one main agent (MA), two league exploiters (LE0, LE1), and one main exploiter (ME) for each side (left or right). The name of each row indicates the agent information as \texttt{Character\_Side\_Checkpoint}. \texttt{Checkpoint=h\_xM} represents a historical version of agent saved at \texttt{x} million steps. The value indicates the win rate of the left (row) player against the right (column) player. For instance, \texttt{ME0\_right} wins all \texttt{MA0\_left\_h\_xM} with high probability, indicating that main exploiters in the league can fully exploit previous main agents. Also the high win rate of \texttt{MA0\_left} against all right agents (except \texttt{MA0\_right}) shows that the main agent at current steps outperforms other agents in the league.}
    \label{fig:league_payoff_main}
\end{figure}

\begin{figure}[t]
% \vskip -0.1in
    \centering
    \includegraphics[width=1\columnwidth]{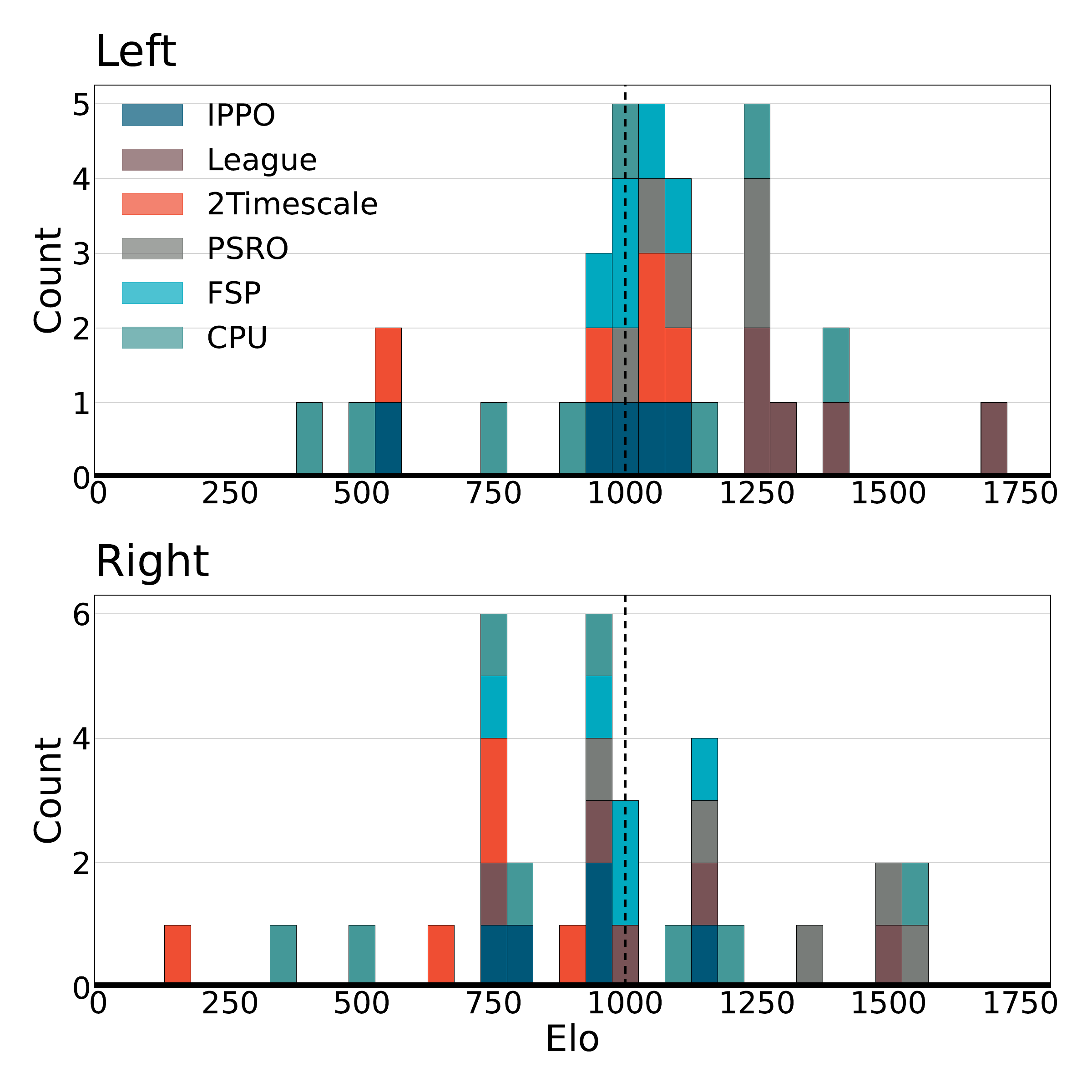}
    \vskip -0.2in
    \caption{The distribution of Elo ratings for top ten agents from each baseline.}
    \label{fig:top_elo}
% \vskip -0.2in
\end{figure}

To answer question \textbf{(b)}, we evaluate five SOTA algorithms mentioned in Section~\ref{sec:baselines}: independent PPO (IPPO), two-timescale IPPO (2Timescale), fictitious self-play (FSP), policy-space response oracles (PSRO), and league training (League) in the scenario \emph{sf\_ryu\_vs\_ryu}. IPPO and 2Timescale can be categorized into the independent learning paradigm, while FSP, PSRO, and League can be categorized into the population-based learning paradigm. For each algorithm, we initialize the population of agents with a pretrained policy in \emph{sf\_ryu\_vs\_ryu(cpu)} against the most difficult CPU\footnote{We do not pre-train in \emph{sf\_ryu\_full\_game} as \emph{sf\_ryu\_vs\_ryu} does not require skills to compete with other characters rather than Ryu.}. We use the transformed actions $\mathcal{A}_\text{trans}$ with hard-coded special moves to unleash the full potential for agents. As a fair comparison, we use the same codebase (\benchname{}-Baselines) and fix the hyperparameters of the backbone PPO algorithm. We train IPPO and 2Timescale for approximately 50M steps until the Elos saturate across all three seeds, FSP and PSRO for approximately 250M steps, and League for approximately 700M steps due to a larger population. A slice of the league during the league training process is visualized in Figure~\ref{fig:league_payoff_main} Please refer to Appendix~\ref{appx:implementation} for more implementation details.

For each algorithm, we report the training Elos of agents in the population during the course of training, respectively. The results are shown in Appendix~\ref{appx:indiv_elo}, which reveal that all baseline algorithms are improving their policies at the onset of training. Subsequently, IPPO and 2Timescale gradually converge and oscillate around the peak Elos, where FSP, PSRO, and League continue to increase their scores. This suggests that IPPO and 2Timescale may suffer from optimization issues during training and population-based methods may be more suitable for policy learning in fighting games.

To compare different baseline algorithms, we select the top ten agents (five on each left or right side) from each algorithm to form a new population, and compute the test Elos for this group of agents and CPU policies. We report the highest Elos for each algorithm in Table~\ref{tab:top_elo} and the distribution of these agents' Elos in Figure~\ref{fig:top_elo}, where we find that League and PSRO significantly outperform other baselines, and population-based methods deliver better results than independent learning counterparts, which is aligned with our previous observation inspecting Elos of baselines individually. On the other hand, we notice that CPU policies may defeat most of the agents in this group except for a few best-performing agents, suggesting that it is still very challenging for existing SOTA algorithms to reach an advanced or superhuman level of performance in these fighting games. We also noticed that two sides of agents reveal asymmetric strengths in terms of Elos in both individual evaluation for each algorithm (Appendix~\ref{appx:indiv_elo} Figure~\ref{fig:ippo_elo}-\ref{fig:league_elo}) and overall evaluations across algorithms (Table~\ref{tab:top_elo}). Such an imbalance may result from various factors, for instance, optimizing instability, variance from the population or Elos computation, etc, and can be an interesting research question for future work.

\begin{table}[]
  % \vskip -0.1in
  \centering
  \caption{Comparison of training steps and the best Elo ratings among baselines, with CPU's Elos as references.}
  \label{tab:top_elo}
  \vskip 0.1in
  % \resizebox{\columnwidth}{!}{
  \begin{tabular}{lcc}
    \toprule
    \textbf{Method} & \textbf{Training Steps} (Left/Right) & \textbf{Elo} (Left/Right) \\
    \midrule
    IPPO & 46M / 46M & 1082 / 1164 \\
    League & 647M / 630M & \textbf{1682} / \textbf{1503} \\
    2Timescale & 51M / 46M & 1080 / 919 \\
    PSRO & 176M / 161M & 1262 / \textbf{1517} \\
    FSP & 262M / 244M & 1079 / 1150 \\
    \textcolor{gray}{CPU} & \textcolor{gray}{N/A} & \textcolor{gray}{1395} / \textcolor{gray}{1541} \\
    \bottomrule
  \end{tabular}
  % }
  % \vskip -0.2in
\end{table}

\subsection{Non-Exploitability of Trained Agents}

\begin{table}[t]
  % \vskip -0.1in
  \centering
  \caption{Comparison of methods' exploitability. A lower number indicates the evaluated policy is more robust to exploitation.}
  \label{tab:exploitability}
  \vskip 0.1in
  \begin{tabular}{lcc}
    \toprule
    \textbf{Method} & \textbf{Exploitability} (Left/Right) \\
    \midrule
    IPPO       & 0.96 $\pm$ 0.03 / 0.91 $\pm$ 0.03 \\
    League     & \textbf{0.94 $\pm$ 0.05} / 0.94 $\pm$ 0.00 \\
    2Timescale & 0.96 $\pm$ 0.02 / 0.90 $\pm$ 0.05 \\
    PSRO       & 0.97 $\pm$ 0.02 / \textbf{0.88 $\pm$ 0.05} \\
    FSP        & 1.00 $\pm$ 0.00 / 0.95 $\pm$ 0.01 \\
    PPO        & 0.99 $\pm$ 0.02 / 0.99 $\pm$ 0.01 \\
    \bottomrule
  \end{tabular}
  % \vskip -0.2in
\end{table}

To answer question \textbf{(c)}, we measure the non-exploitability of baseline algorithms according to the evaluation approaches proposed in Section~\ref{sec:metrics}. More specifically, we choose models with the highest Elos from each two-player baseline algorithm respectively, and compare their exploitability with the single-player pretrained model used for initializing the population-based methods in Section~\ref{sec:2p_exp}.

The practical exploitability is calculated by setting the trained policy fixed on one side, and deploying a PPO agent on the other side as an exploiter. The PPO exploiter will be trained until convergence, and the success rate of the exploiter is the estimated exploitability of the original policy, according to Definition~\ref{def:exploitability}.

\paragraph{Single-Agent RL Exploiters.} 
We use PPO as the algorithm for training exploiters, given its decent performance in both single-player and two-player scenarios shown in previous experiments. Table~\ref{tab:exploitability} shows the exploitability of comparing methods evaluated across three seeds, from which we observe that the single-player pretrained policy via PPO is easier to exploit and suffers from higher exploitability than almost all selected policies from two-player baselines. Therefore, this result indicates that two-player learning algorithms such as League and PSRO can help to improve the robustness of learned policies. On the other hand, the PPO exploiter eventually learns to beat policies from all baselines (with a win rate greater than 0.5), which means that none of these algorithms can result in the exact Nash equilibrium policies, or even close to it. Therefore, closing this gap is a challenging direction for future research.

\paragraph{Human Players as Exploiters.}
In addition to exploiting the learned models with RL algorithms, we also attempt to exploit their policies with human effort. During human evaluations, the evaluated models reveal some robustness to human players (e.g., defend when a human player attacks), but some simple strategies (e.g., defensive posture combined with low kicks at proper timing) could still defeat them rather consistently. Visualizations are provided in Appendix~\ref{appx:human}. 

Therefore, based on two exploiting experiments, we observe that \textit{existing competitive MARL algorithms are found hard to learn non-exploitable strategies in competitive fighting games like Street Fighter}, thus raising a new challenge for the research community.

\section{Conclusion and Limitation}
In this paper, we present the \benchname{} platform and evaluation benchmarks as a novel testbed for competitive MARL research. The platform supports various video action games including the popular Street Fighter series, with flexible support for new game integration. 

We further provide experimental evaluations of present RL and MARL algorithms in both single-player and two-player modes of one specific game Street Fighter. In the single-player setting, we proposed a learning scheme based on curriculum learning. It trains a general RL agent that can consistently beat CPUs across different characters. In the two-player setting, the Elo rating and exploitability test are conducted as part of the proposed evaluation criteria. Our implementation of league training and PSRO provides stronger agents than FSP and IPPO in terms of Elo ratings. However, both single-agent RL and human players are capable of exploiting all agents learned by current widely adopted algorithms. 

Our current work is limited to fully competitive two-player games. One important challenge of MARL is its diverse nature, which includes collaborative games, competitive games, two-player games, and multiplayer games, all of which have rather different problem structures and solution concepts. The more general setting, which involves more than two players and both cooperation and competition, is not
yet explored and should be an important future direction. Although FightLadder supports multiple fighting games, our current results are mostly conducted on Street Fighter, and we are curious to see more results on other games.

This work motivates further research in developing more efficient and effective self-play algorithms finding non-exploitable strategies. We hope that our platform prompts general interest and more extensive research in competitive MARL and serves as a standard benchmark for developing practically useful self-play training paradigms.

\clearpage
% Acknowledgements should only appear in the accepted version.
\section*{Acknowledgements}
This work was supported by Office of Naval Research N00014-22-1-2253, National Science Foundation Grant NSF-IIS-2107304, and  National Science Foundation Graduate Research Fellowship Program under Grant No. DGE-2039656.

\section*{Impact Statement}
This work may advance the field of game AI, thus has potentials to affect the gaming experience for human players. The strong AI agents for popular fighting games may attract people's attention to get involved in these games, or make them feel that the games can be even more challenging for human. Another positive impact is that our study promotes the research for robust systems against adversarial attacks.

% \clearpage
% In the unusual situation where you want a paper to appear in the
% references without citing it in the main text, use \nocite
% \nocite{langley00}

\bibliography{main}
\bibliographystyle{icml2024}

%%%%%%%%%%%%%%%%%%%%%%%%%%%%%%%%%%%%%%%%%%%%%%%%%%%%%%%%%%%%%%%%%%%%%%%%%%%%%%%
%%%%%%%%%%%%%%%%%%%%%%%%%%%%%%%%%%%%%%%%%%%%%%%%%%%%%%%%%%%%%%%%%%%%%%%%%%%%%%%
% APPENDIX
%%%%%%%%%%%%%%%%%%%%%%%%%%%%%%%%%%%%%%%%%%%%%%%%%%%%%%%%%%%%%%%%%%%%%%%%%%%%%%%
%%%%%%%%%%%%%%%%%%%%%%%%%%%%%%%%%%%%%%%%%%%%%%%%%%%%%%%%%%%%%%%%%%%%%%%%%%%%%%%
\newpage
\appendix
\onecolumn
% \section{You \emph{can} have an appendix here.}

% You can have as much text here as you want. The main body must be at most $8$ pages long.
% For the final version, one more page can be added.
% If you want, you can use an appendix like this one.  

% The $\mathtt{\backslash onecolumn}$ command above can be kept in place if you prefer a one-column appendix, or can be removed if you prefer a two-column appendix.  Apart from this possible change, the style (font size, spacing, margins, page numbering, etc.) should be kept the same as the main body.

\section{Details of FightLadder}

\subsection{Dense Reward}
\label{appx:dense_reward}
The shaped dense reward for the $i$-th agent at step $t$ is defined as follows:
\begin{align}
    r_{i,t} = \alpha\left[\lambda(\text{HP}_{-i,t-1}-\text{HP}_{-i,t})-(\text{HP}_{i,t-1}-\text{HP}_{i,t}) + r_{i,\text{bonus}}\right],
\end{align}
where $\alpha$ is a scaling factor, $\text{HP}_{i,t}$ denotes agent $i$'s hit-point at step $t$ and $\lambda$ control the aggressiveness of learned agents, and $-i$ denotes the opponent agent. At the end of the game, the agent $i$ will receive a bonus reward $r_{i,\text{bonus}}$, which is positively correlated to $\text{HP}_i$ if it wins and negatively correlated to $\text{HP}_{-i}$ if it loses. By default, we choose $\lambda=3$ in SF2, FF2, and MK, and $\lambda=1$ in SF3 and KOF97, for the consideration of practical performances.

\subsection{Game Settings} \label{appx:scenarios}
Table~\ref{tab:games} illustrates the observation, action, and rewards as well as other elements in the environment for all supported games --- Street Fighter II (SF2), Fatal Fury 2 (FF2), Mortal Kombat (MK), Street Fighter III (SF3), and The King of Fighters '97 (KOF97).

\begin{table}[!h]
  \vskip -0.1in
  \centering
  \caption{Specification of supported games in \benchname{}.}
  \label{tab:games}
  \vskip 0.1in
  \small
  \resizebox{\textwidth}{!}{
  \begin{tabular}{lccccc}
    \toprule
    & \textbf{SF2} & \textbf{FF2} & \textbf{MK} & \textbf{SF3} & \textbf{KOF97} \\
    \midrule
    Observation (Pixels) & 100$\times$128$\times$3 & 112$\times$128$\times$3 & 112$\times$160$\times$3 & 112$\times$192$\times$3 & 112$\times$192$\times$3 \\
    Human Action Supported & Yes & Yes & Yes & Yes & Yes \\
    Transformed Action Supported & Yes & Yes & Yes & No & No \\
    Shaped Dense Reward & Yes & Yes & Yes & Yes & Yes  \\
    Default Frames Per Step & 8 & 8 & 8 & 3 & 3 \\
    Default Frames Stacked\tablefootnote{We uniformly sample the stacked frames as observations to improve the computational efficiency.} & 12 & 12 & 12 & 9 & 9 \\
    \multirow{2}{*}{Additional Available Info} & HPs, Countdown, & HPs, Countdown & HPs, Countdown, & HPs & HPs, Countdown, \\ 
    & Scoreboard, Positions & & Scoreboard & & Positions, Power Status \\
    \bottomrule
  \end{tabular}
  }
  % \vskip -0.2in
\end{table}

\subsection{Comparison of MARL Game Platforms} \label{appx:compare_platforms}
Table~\ref{tab:environment_comparison} compares our FightLadder with several popular MARL game platforms mostly focusing on competitive settings, in terms of observation space, action space, whether baseline methods are included and the number of agents in games. For the observation space, `Continuous' indicates a vector-form latent state information of the game with continuous numerical values, and `Image' indicates visual RGB information as observations. PommerMan~\citep{resnick2018pommerman} uses grid environments therefore its observation only has discrete values. For the action space, most of the games only involves discrete action values except for Arena~\citep{song2020arena}. For the number of agents in these platforms, MPE provide diverse competitive settings like 1v1, 1v$N$, 1v1v1 and so on. MAgent includes 1 million agents competing againts each other, and for Neural MMO~\citep{suarez2021neural} the number of agents is 256 or 1024. The team mode in our FightLadder and Arena supports the competitive settings of two teams, where each team includes multiple characters to be controlled by one team policy or separate agent policies.

\begin{table}[h!]
\centering
\caption{Comparison of popular MARL game platforms.}
\vskip 0.1in
\resizebox{\textwidth}{!}{
\begin{tabular}{ccccc}
\hline
\textbf{Env} & \textbf{Observation Space} & \textbf{Action Space} & \textbf{Baselines} & \textbf{\# Agents} \\ \hline
MPE \citep{mordatch2018emergence} & Continuous & Discrete & Yes & 1v1, 1v$N$ and 1v1v1... \\ \hline
MAgent \citep{zheng2018magent} & Continuous+Image & Discrete & Yes & 1 million \\ \hline
Arena \citep{song2020arena} & Continuous+Image & Continuous/Discrete & Yes & 1v1, $N$v$N$ and team mode \\ \hline
Neural MMO \citep{suarez2021neural} & Continuous & Discrete & Yes & 256 and 1024 \\ \hline
PettingZoo Atari \citep{terry2021pettingzoo} & Continuous+Image & Discrete & No & 1v1 \\ \hline
PommerMan \citep{resnick2018pommerman} & Discrete & Discrete & No & 2v2 \\ \hline
FightLadder (Ours) & Continuous+Image & Discrete & Yes & 1v1 and team mode \\ \hline
\end{tabular}
}
\label{tab:environment_comparison}
\end{table}

\section{Baseline Algorithms of \benchname{}-Baselines} \label{appx:baselines}
\paragraph{Independent Learning (IPPO).} 
Independent learning is a straightforward extension of single-agent RL into MARL. It decomposes the joint optimization into individual ones for each agent while regarding all other agents as part of the environment. It can be implemented easily by simultaneously running single-agent RL algorithms for each player.
Theoretically, this independent learning paradigm suffers from suboptimality~\citep{tan1993multi, foerster2018counterfactual}, because the environment becomes non-stationary while other agents are updating their policies.  
However, recent work~\citep{de2020independent, yu2022surprising} finds that with modest hyperparameter tuning, IPPO can serve as a strong baseline compared to other state-of-the-art algorithms in some cooperative MARL tasks. 

\paragraph{Two-timescale Learning (2Timescale).}
Two-timescale learning follows the independent learning paradigm, but requires two players to update gradients according to the two-timescale rule, i.e., one player uses a much smaller step size than the other one.
As a result of this modification, two-timescale learning enjoys some nice theoretical properties --- it is proven that under some mild assumptions, independent policy gradient algorithms satisfying two-timescale converge to a Nash equilibrium in two-player zero-sum stochastic games~\citep{daskalakis2020independent}.

\paragraph{Population-Based Methods.}
The independent learning framework is only training agents against the current version of their opponents, which may fail or converge slowly due to the lack of diversity~\citep{dresher2016advances}. Population-based methods are proposed to increase policy diversity by maintaining a pool of policies in previous iterations, and using them as a curriculum to update the current policy. More specifically, for $t$-th update, the agent $\mu^t$ plays with previous versions of its opponent $\Tilde{\nu}$ sampled from the meta-strategy $\rho_\nu$, which is a distribution over $\nu^0,\nu^1,\dots,\nu^{t-1}$. Algorithm~\ref{alg:population} presents the pseudo-code for general population-based methods. With different choices of sampling distribution, we can recover several state-of-the-art baselines:
\begin{itemize}
    \item \textbf{Fictitious Self-Play (FSP),} where $\rho_\nu$ is the uniform distribution~\citep{pmlr-v37-heinrich15}: $\text{Uniform}(\nu^0,\nu^1,\dots,\nu^{t-1})$.
    \item \textbf{Policy-Space Response Oracles (PSRO),} where $(\Tilde{\mu}, \Tilde{\nu})$ are sampled from the meta-strategy $(\rho_\mu, \rho_\nu)$ by solving Nash equilibrium of the payoff matrix game between $\mu^0,\mu^1,\dots,\mu^{t-1}$ and $\nu^0,\nu^1,\dots,\nu^{t-1}$~\citep{lanctot2017unified}.
    \item \textbf{League Training (League),} where three types of agents --- main agents, league exploiters, and main exploiters, are introduced into the population. Main agents train against themselves as well as all previous versions of agents in the population; league exploiters train against all previous agents; and main exploiters optimize the best response of main agents. Each type of agent adopts a different sampling distribution which is a mixture of self-play and prioritized fictitious self-play. We refer readers to~\citep{vinyals2019grandmaster} for more implementation details.
\end{itemize}

\begin{algorithm}[!h]
\caption{Population-Based Methods for MGs}
\label{alg:population}
\begin{algorithmic}[1]
\STATE Initialize policies $\mu^0=\{\mu_h\}, \nu^0=\{\nu_h\}, h\in[H]$
\STATE Initialize policy sets: $\mu=\{\mu^0\}, \nu=\{\nu^0\}$
\STATE Initialize meta-strategies: $\rho_\mu=[1.], \rho_\nu=[1.]$

\FOR{$t=1,\ldots,T$}
    \IF{$t\%2==0$}
     \STATE  $\nu^t = \textsc{Best\_Response}(\rho_\mu, \mu)$
     \STATE $\nu=\nu\bigcup\{\nu^t\}$
     \STATE Update $\rho_\nu$ according to specific algorithms
    \ELSE
    \STATE  $\mu^t = \textsc{Best\_Response}(\rho_\nu, \nu)$
    \STATE $\mu=\mu\bigcup\{\mu^t\}$
    \STATE Update $\rho_\mu$ according to specific algorithms
    \ENDIF
    % \STATE exploitability = $\textcolor{magenta}{\textsc{Best\_Response\_Value}}(\rho_\mu,\mu)+\textcolor{magenta}{\textsc{Best\_Response\_Value}}(\rho_\nu,\nu)$
\ENDFOR
\STATE Return $\mu, \rho_\mu, \nu, \rho_\nu$
\end{algorithmic}
\end{algorithm}

\section{Experiment Details} \label{appx:implementation}

\subsection{Hyperparameters (Table~\ref{tab:ppo_hyperparam} and~\ref{tab:mf_hyperparam})}

\begin{table}[!h]
\centering
% \resizebox{\columnwidth}{!}{
\begin{tabular}{r|l}
\toprule
\textbf{Hyperparameters} & \textbf{Value} \\ \midrule
feature extractor & CNN~\citep{mnih2015human} \\
rollout steps for each environment & 512 \\
batch size & 1024 \\
epochs per update & 4 \\ 
$\gamma$ & 0.94 \\
GAE $\lambda$ & 0.95 \\
learning rate & linear schedule from 2.5e-4 to 2.5e-6 \\
clipping range & linear schedule from 0.15 to 0.025 \\
advantage normalization & True \\
entropy coefficient & 0.0 \\
gradient clipping & 0.5 \\
value function coefficient & 0.5 \\
\bottomrule
\end{tabular}
% }
 \caption{Training hyperparameters for PPO, which is the backbone for both single-player and two-player algorithms in the experiment.}
\label{tab:ppo_hyperparam}
\end{table}

\begin{table}[!h]
\centering
\resizebox{\columnwidth}{!}{
\begin{tabular}{ll|ll|ll}
\toprule
\multicolumn{2}{c|}{\textbf{FSP}} & \multicolumn{2}{c|}{\textbf{PSRO}} & \multicolumn{2}{c}{\textbf{League}} \\ \midrule
\# envs per learner & 24 & \# envs per learner & 24 & \# envs per learner & 24 \\
steps for BR & 10M & steps for BR & 10M & steps for BR & 10M \\
total steps & 50M & total steps & 250M & total steps & 700M \\
\# main agent & 1 & \# main agent & 1 & \# main agent & 1 \\
 &  & Nash solver & ECOS & \# main exploiter & 1 \\ 
 &  &  & \citep{Domahidi2013ecos} & \# league exploiter & 2 \\
\bottomrule
\end{tabular}
}
 \caption{Training hyperparameters for FSP, PSRO, and League. We omit the details of League's opponent scheduling here as it strictly follows the pseudocode provided in~\citep{vinyals2019grandmaster}.}
\label{tab:mf_hyperparam}
\end{table}

\subsection{Training Details}
\label{subsec:app_train_detail}
\begin{figure}[!h]
    \centering
    \includegraphics[width=0.49\textwidth]{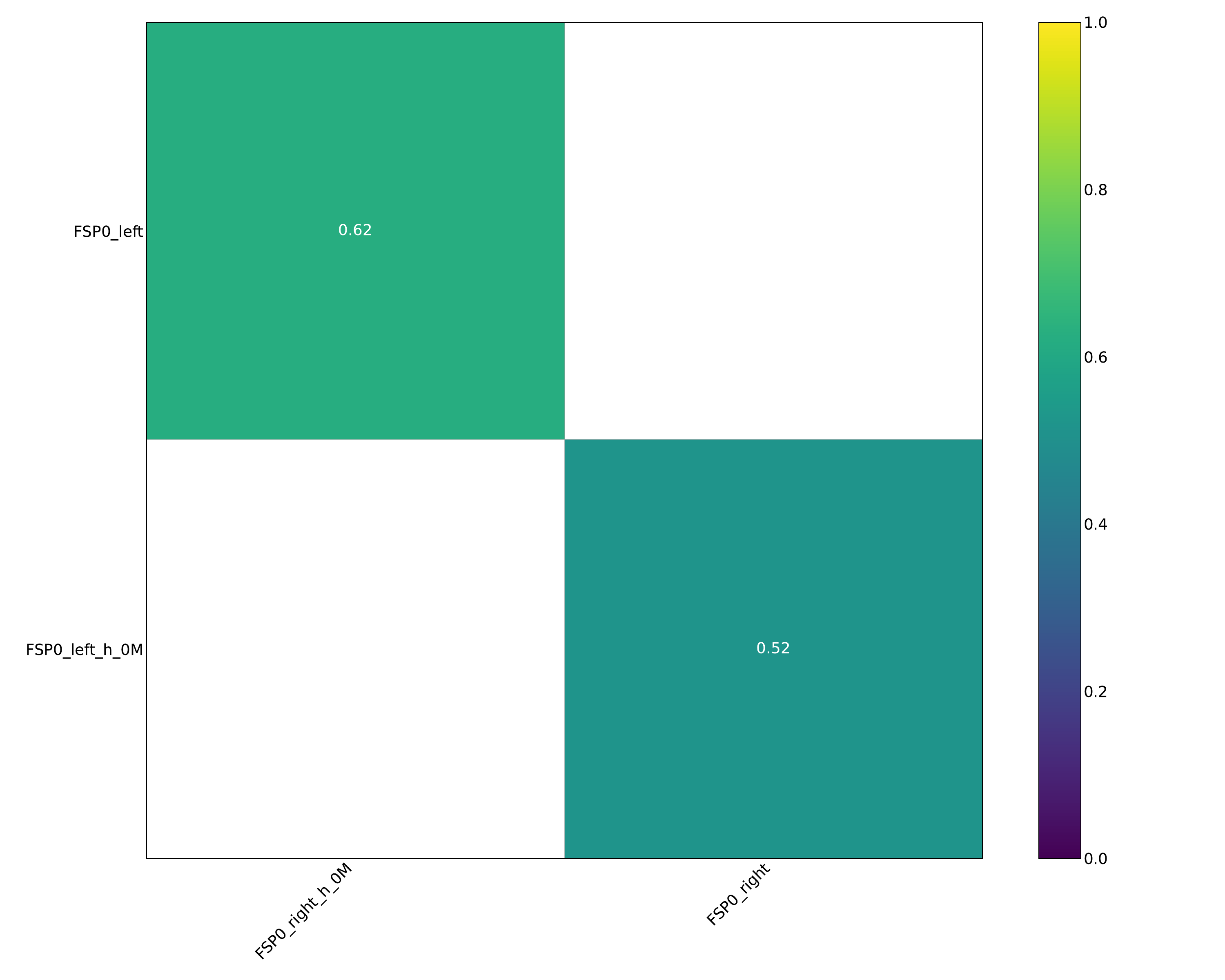}
    \includegraphics[width=0.49\textwidth]{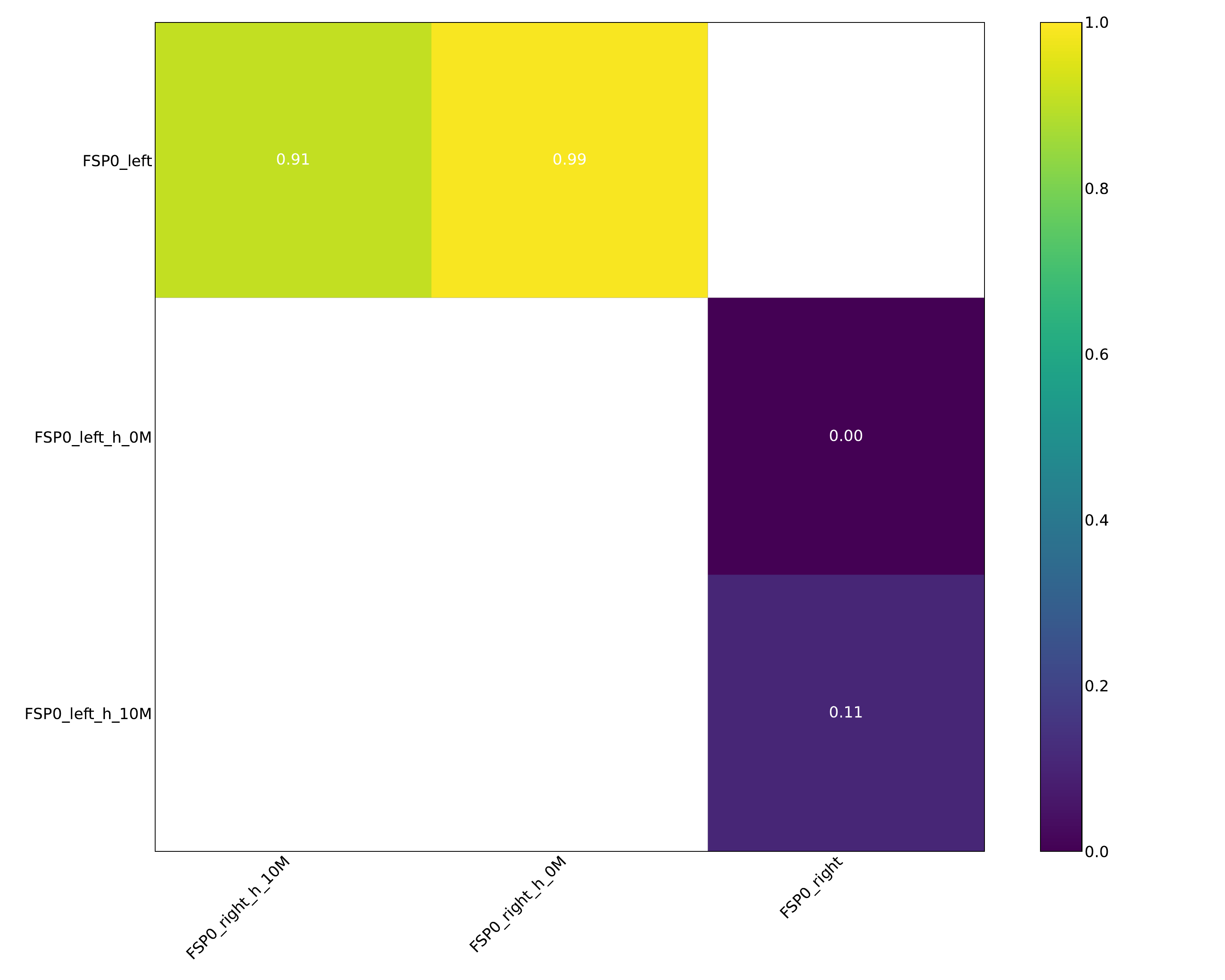}
    \includegraphics[width=0.49\textwidth]{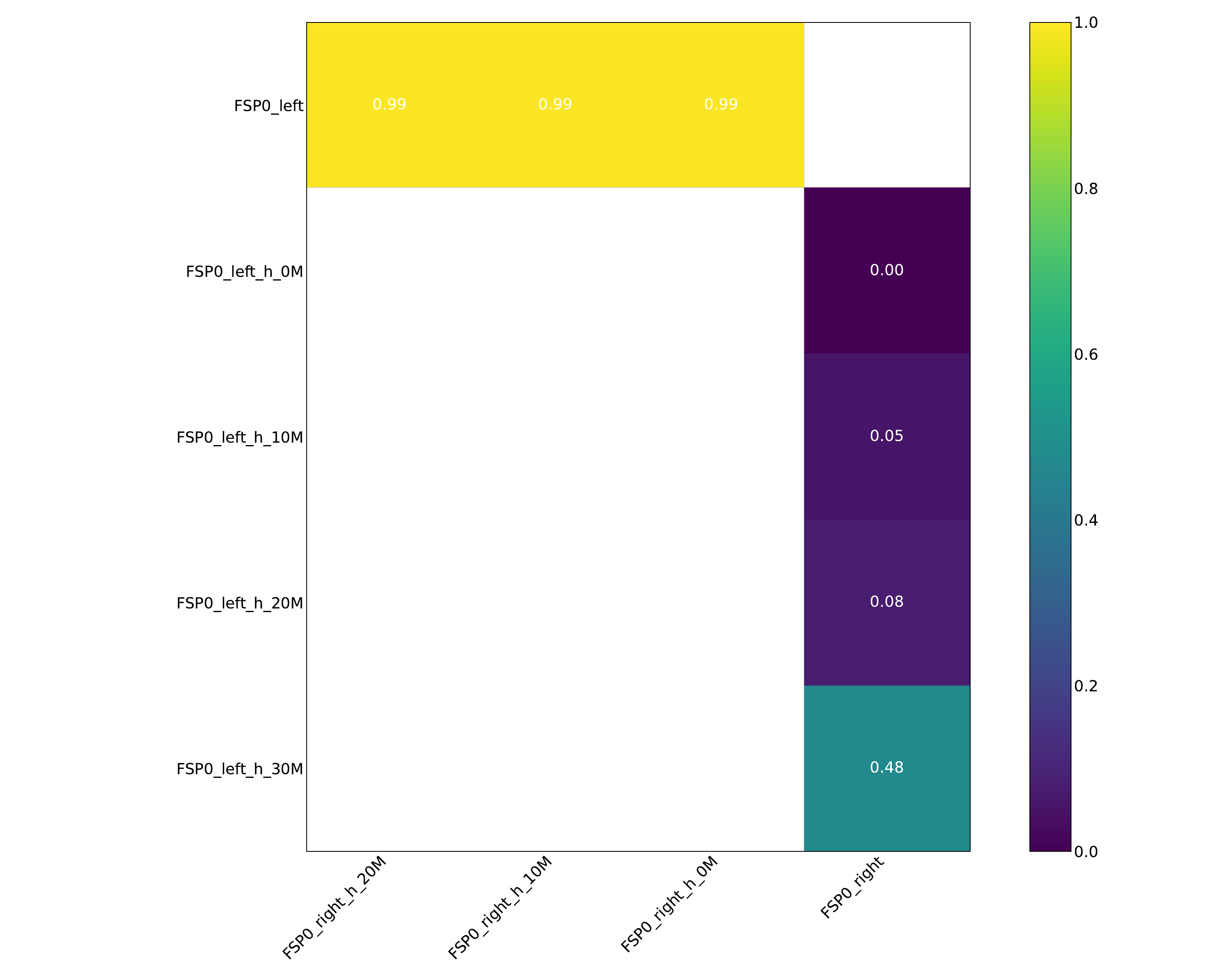}
    \includegraphics[width=0.49\textwidth]{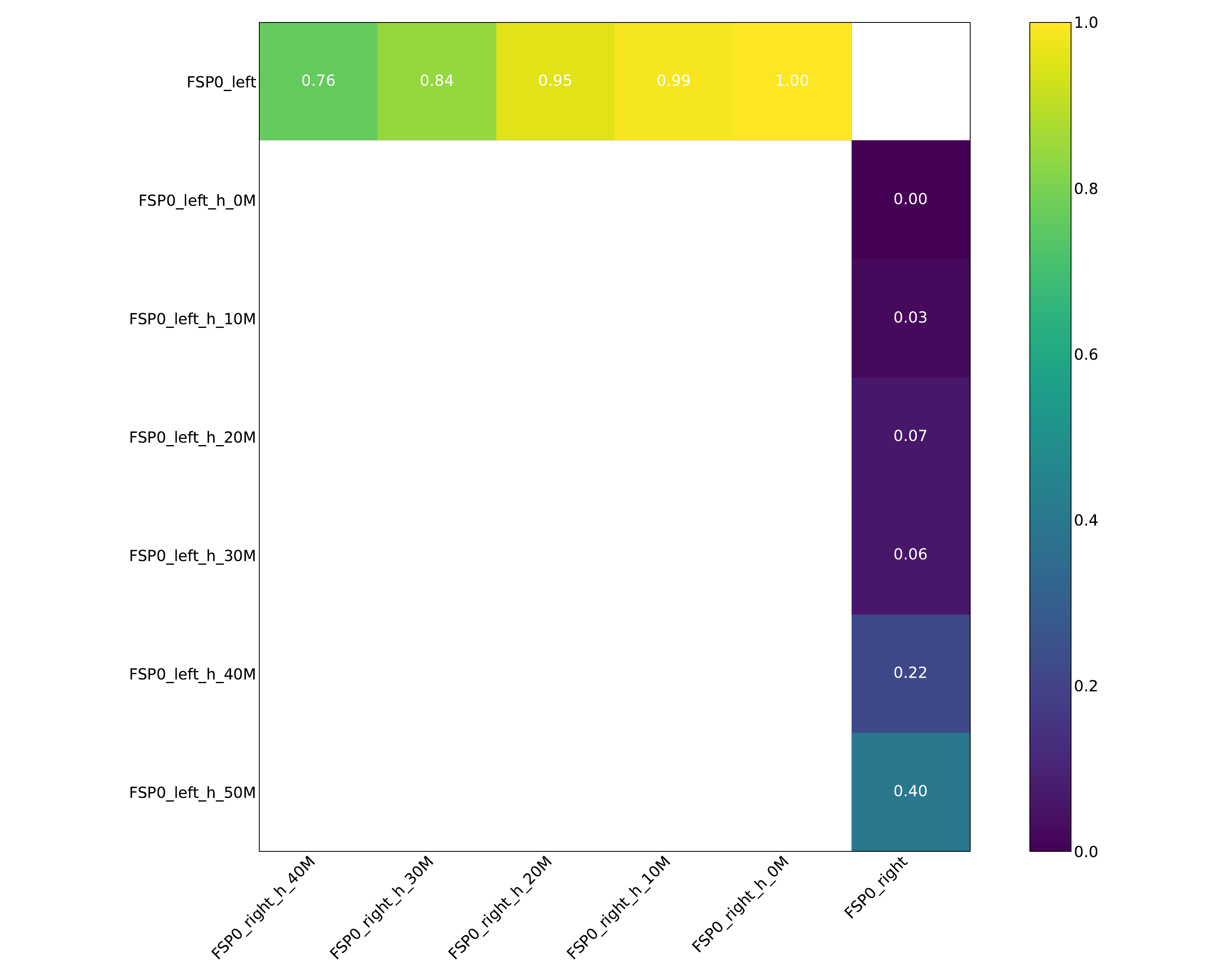}
    \includegraphics[width=0.49\textwidth]{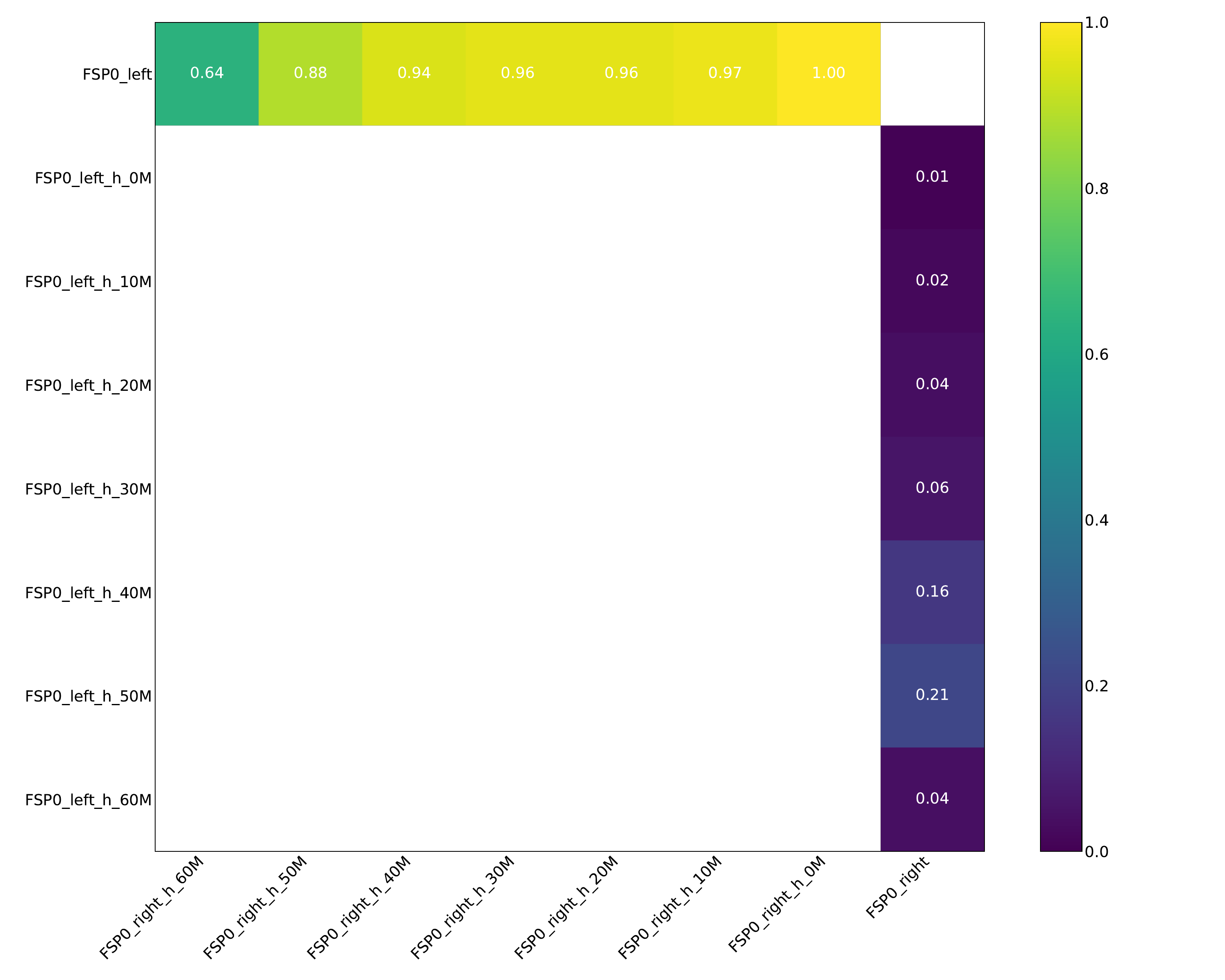}
    \includegraphics[width=0.49\textwidth]{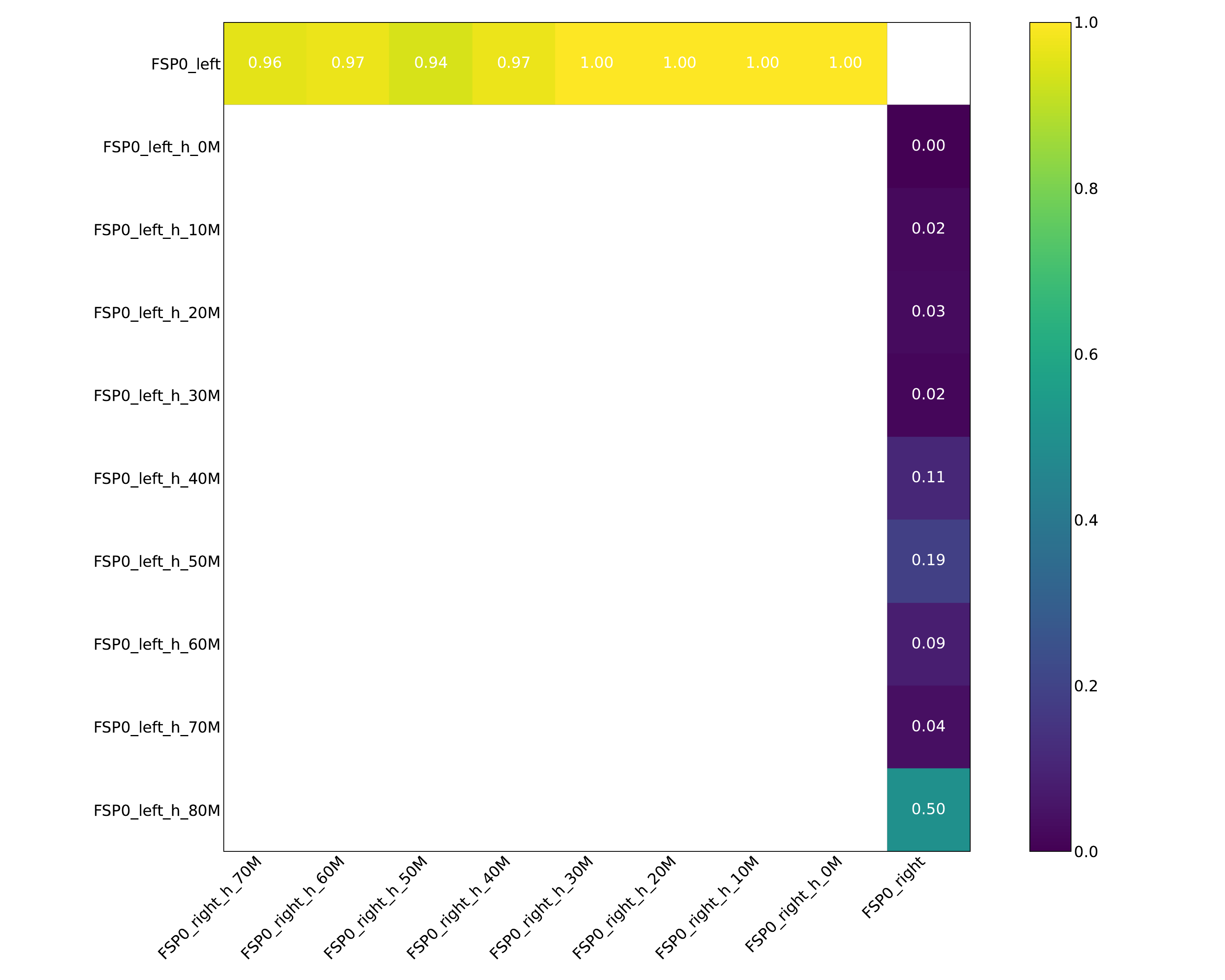}
    \caption{FSP details (training order from top left to bottom right): For FSP, there is one agent for each side (left or right). The name of each row indicates the agent information as \texttt{Character\_Side\_Checkpoint}. \texttt{Checkpoint=h\_xM} represents a previous version of agent saved at \texttt{x} million steps. The value indicates the win rate of the left (row) player against the right (column) player.}
    \label{fig:fsp_payoff}
\end{figure}

\begin{figure}[!h]
    \centering
    \includegraphics[width=0.49\textwidth]{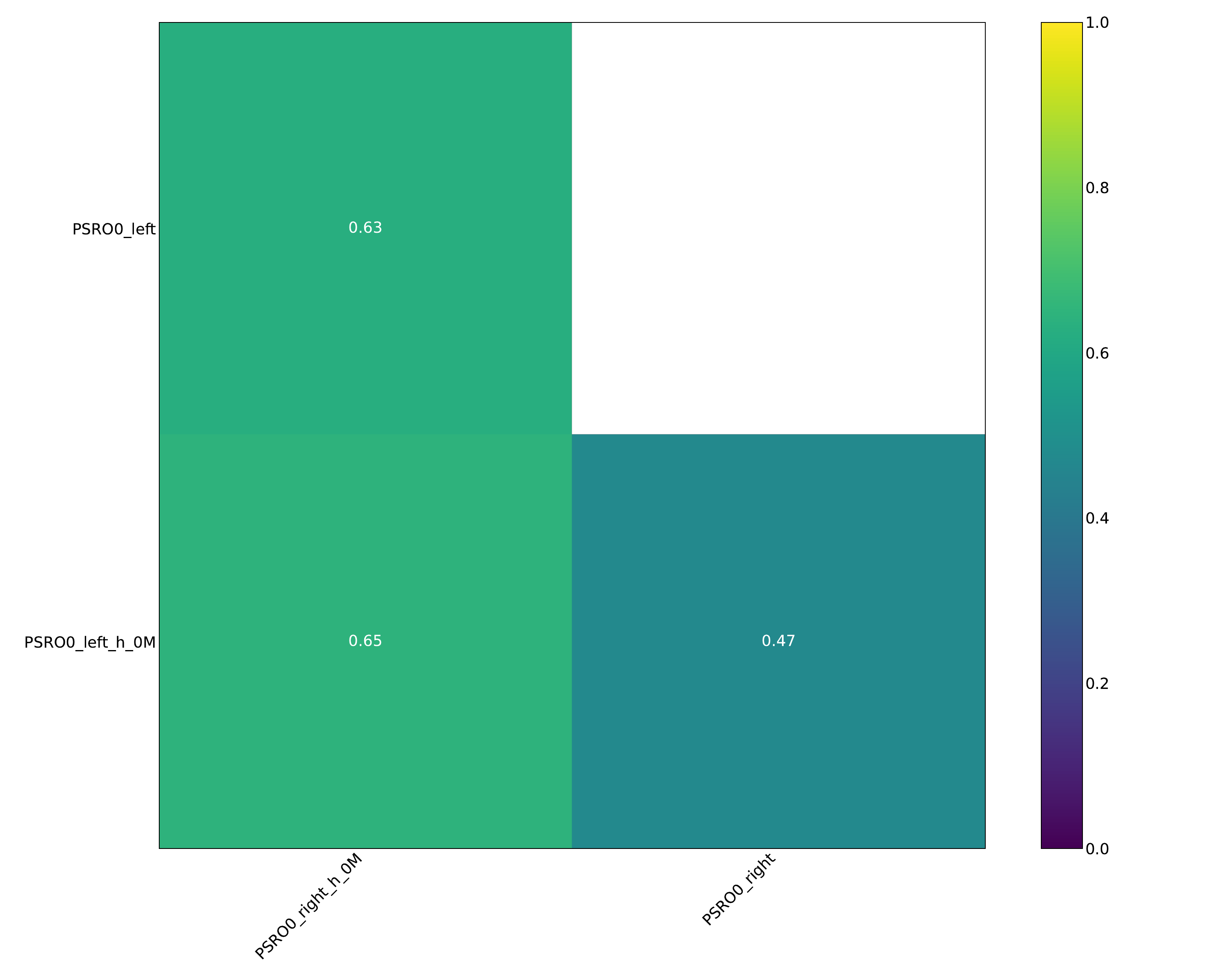}
    \includegraphics[width=0.49\textwidth]{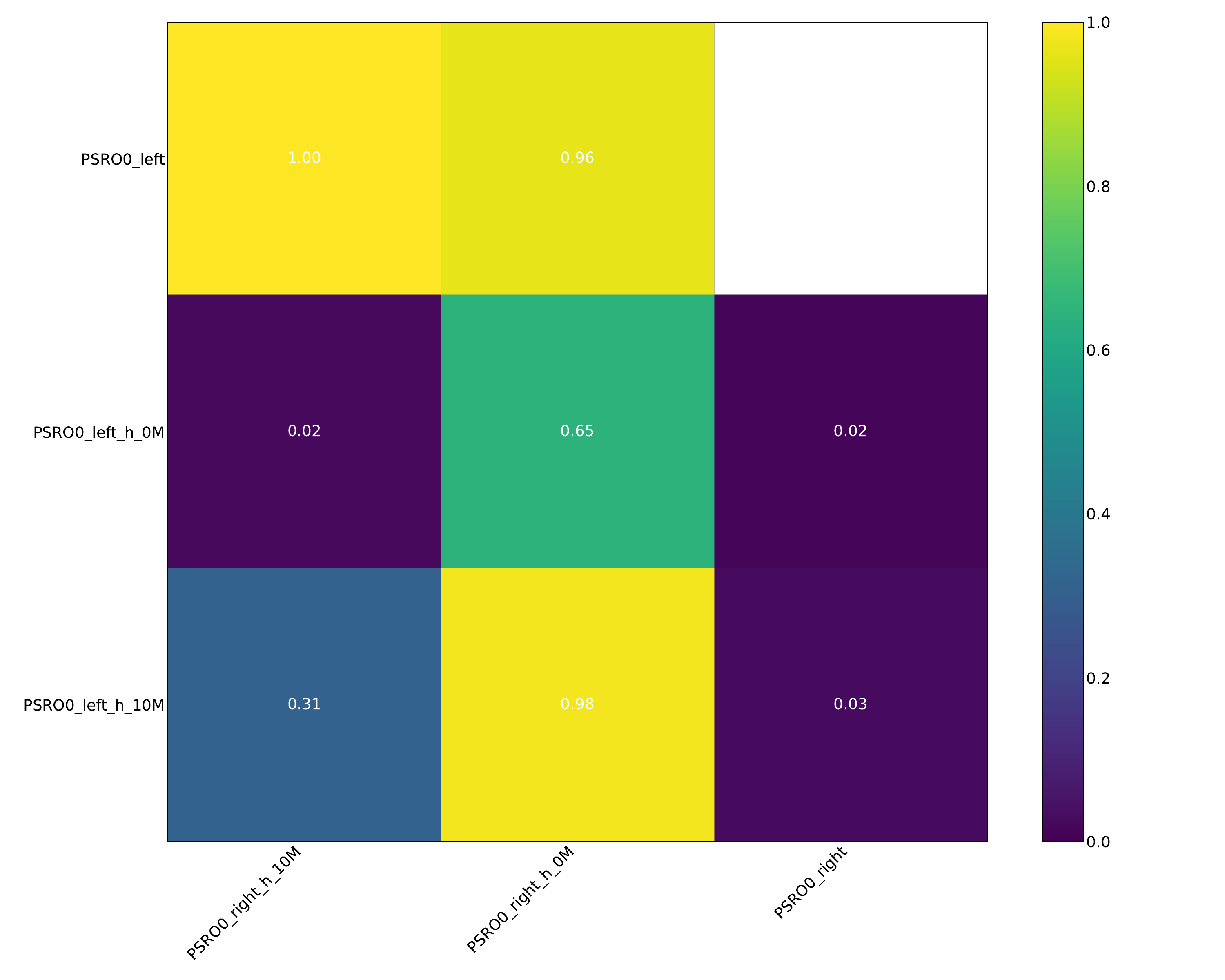}
    \includegraphics[width=0.49\textwidth]{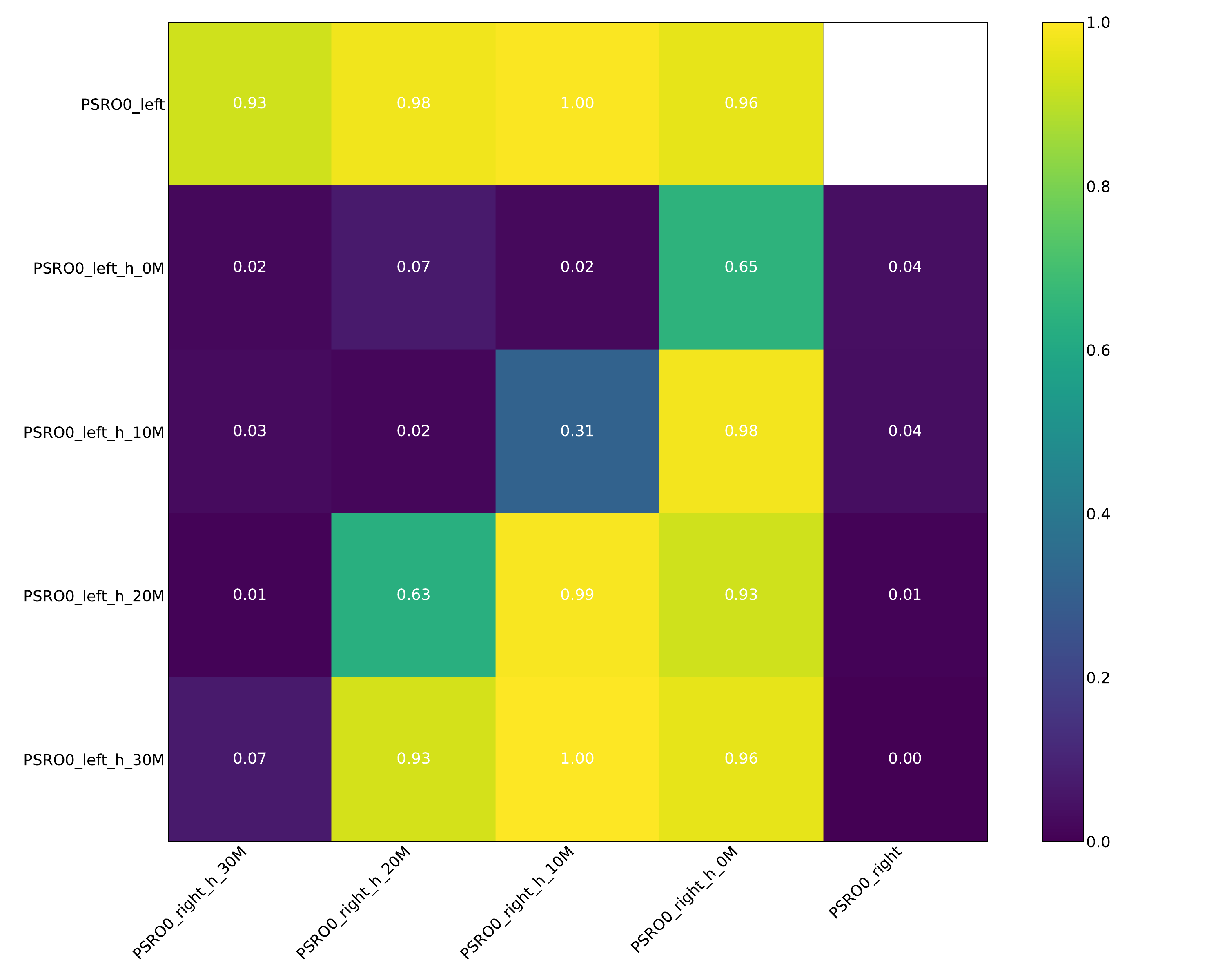}
    \includegraphics[width=0.49\textwidth]{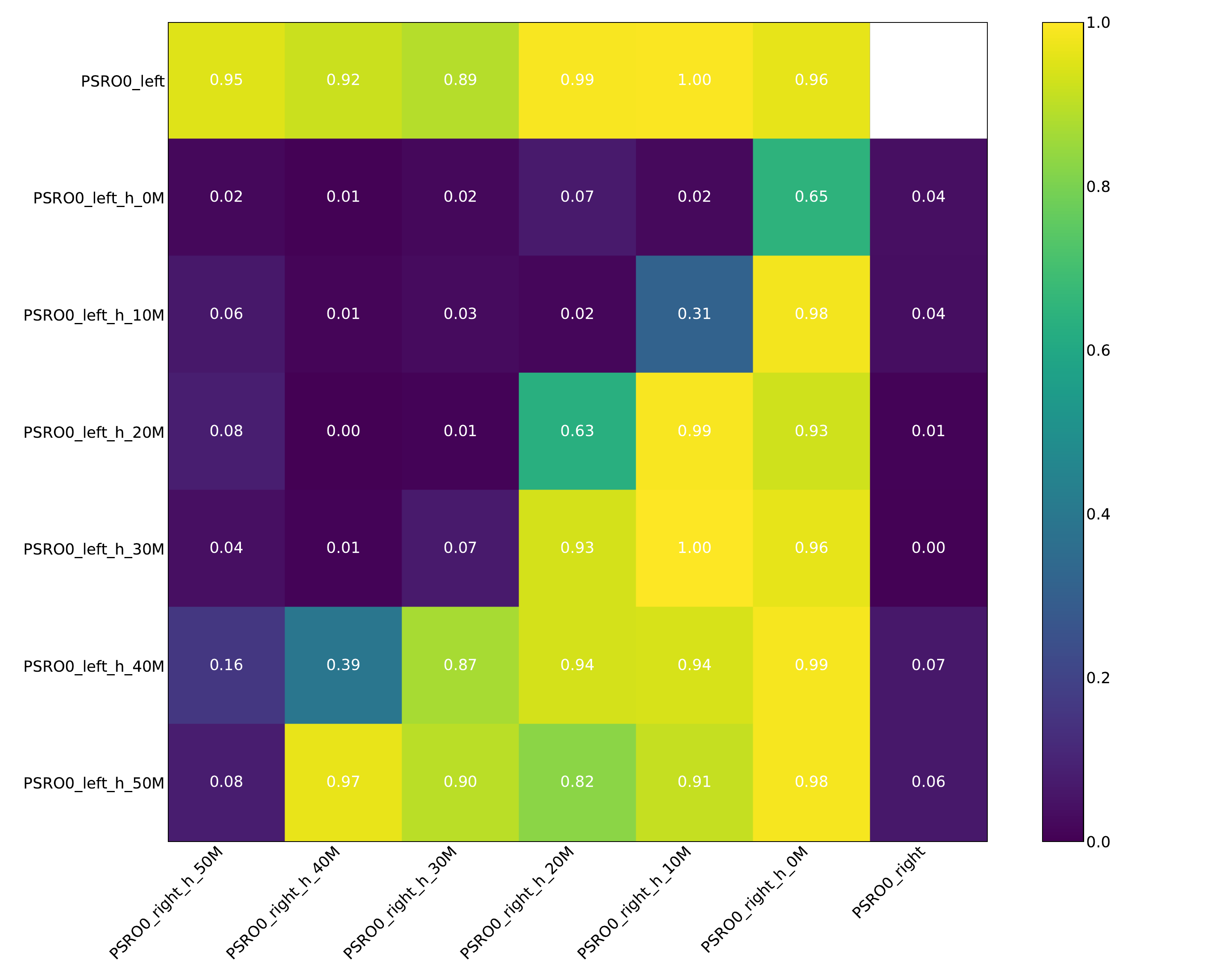}
    \includegraphics[width=0.49\textwidth]{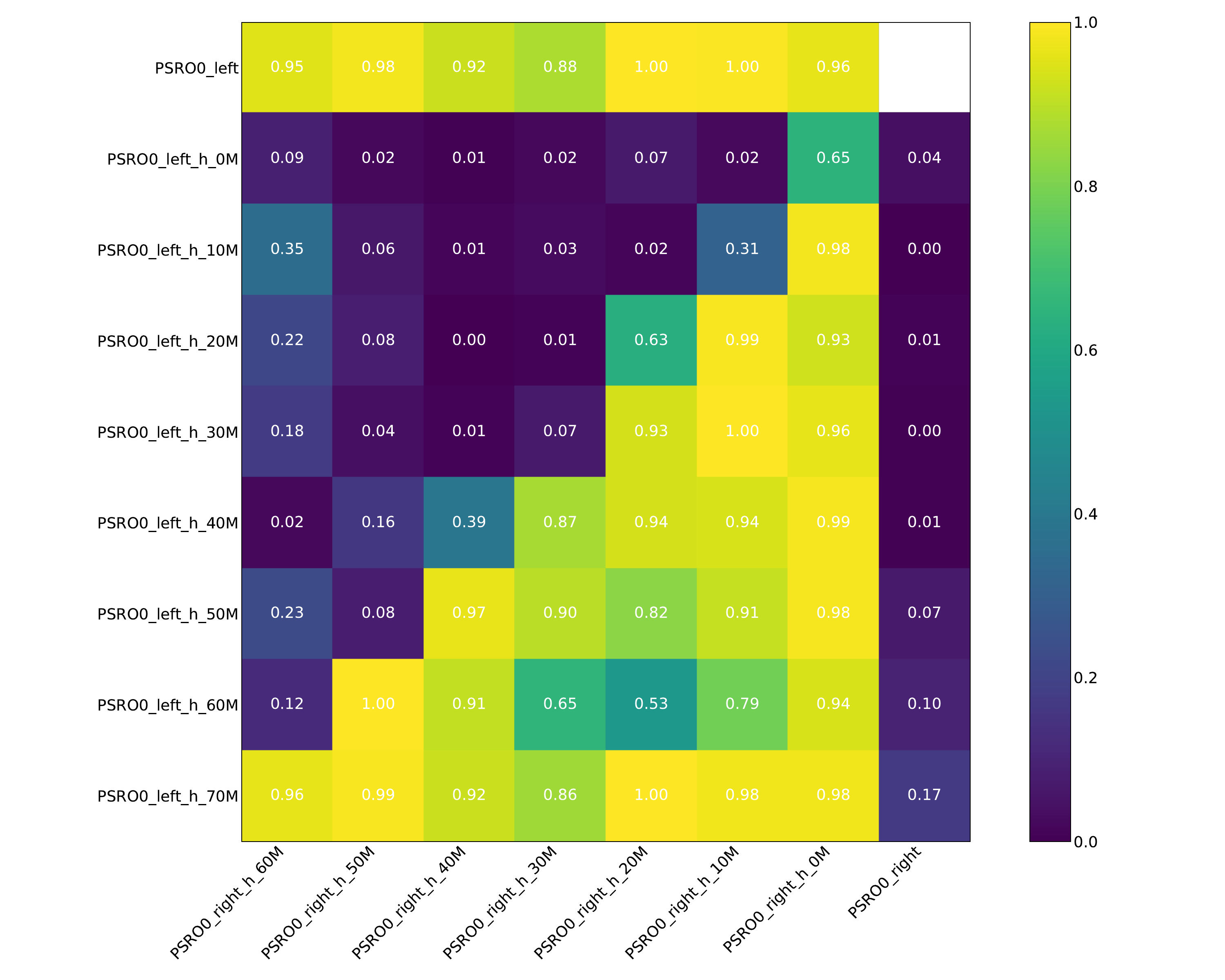}
    \includegraphics[width=0.49\textwidth]{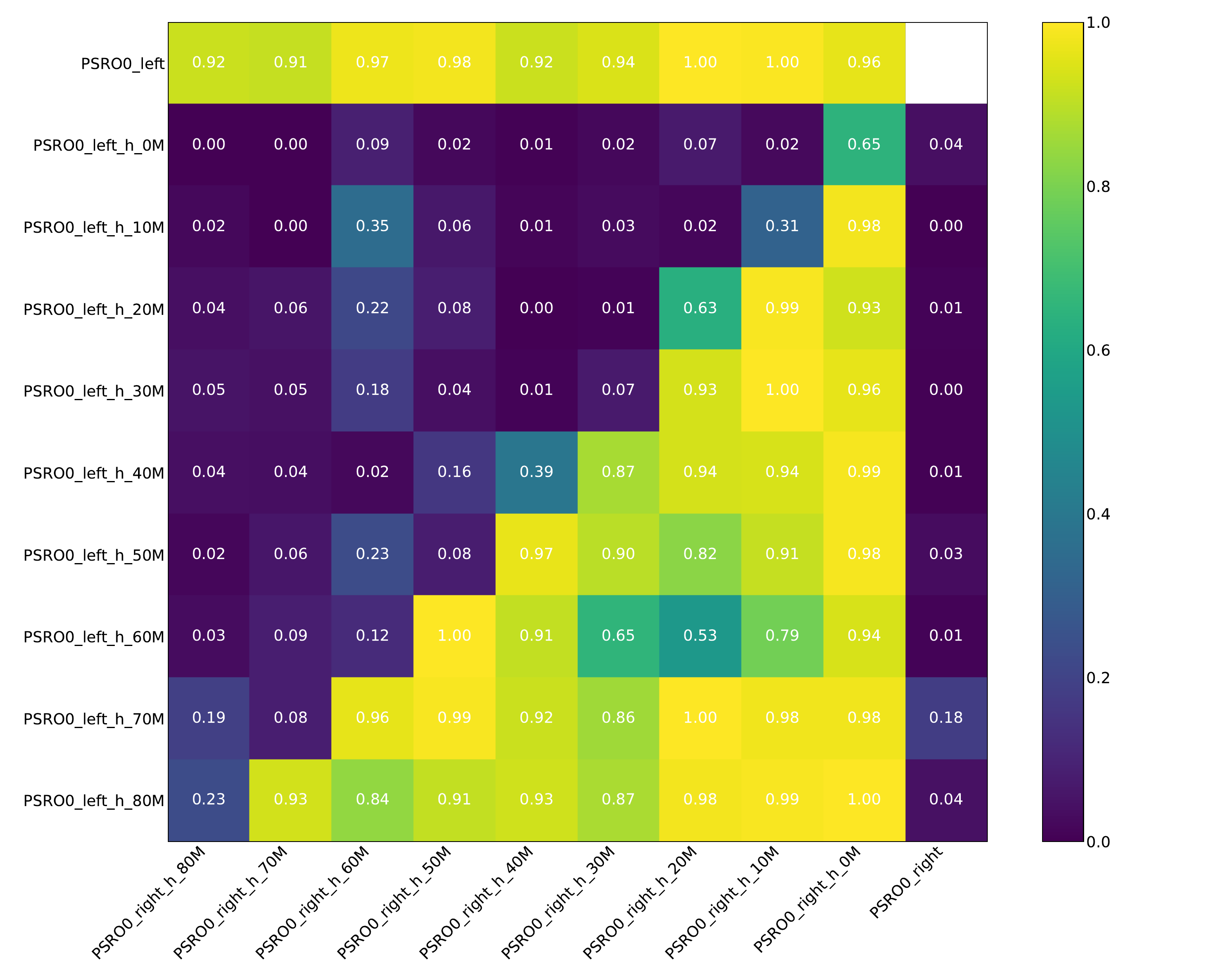}
    \caption{PSRO details (training order from top left to bottom right): For PSRO, there is one agent for each side (left or right). The name of each row indicates the agent information as \texttt{Character\_Side\_Checkpoint}. \texttt{Checkpoint=h\_xM} represents a previous version of agent saved at \texttt{x} million steps. The value indicates the win rate of the left (row) player against the right (column) player.}
    \label{fig:psro_payoff}
\end{figure}

\begin{figure}[!h]
    \centering
    \includegraphics[width=0.49\textwidth]{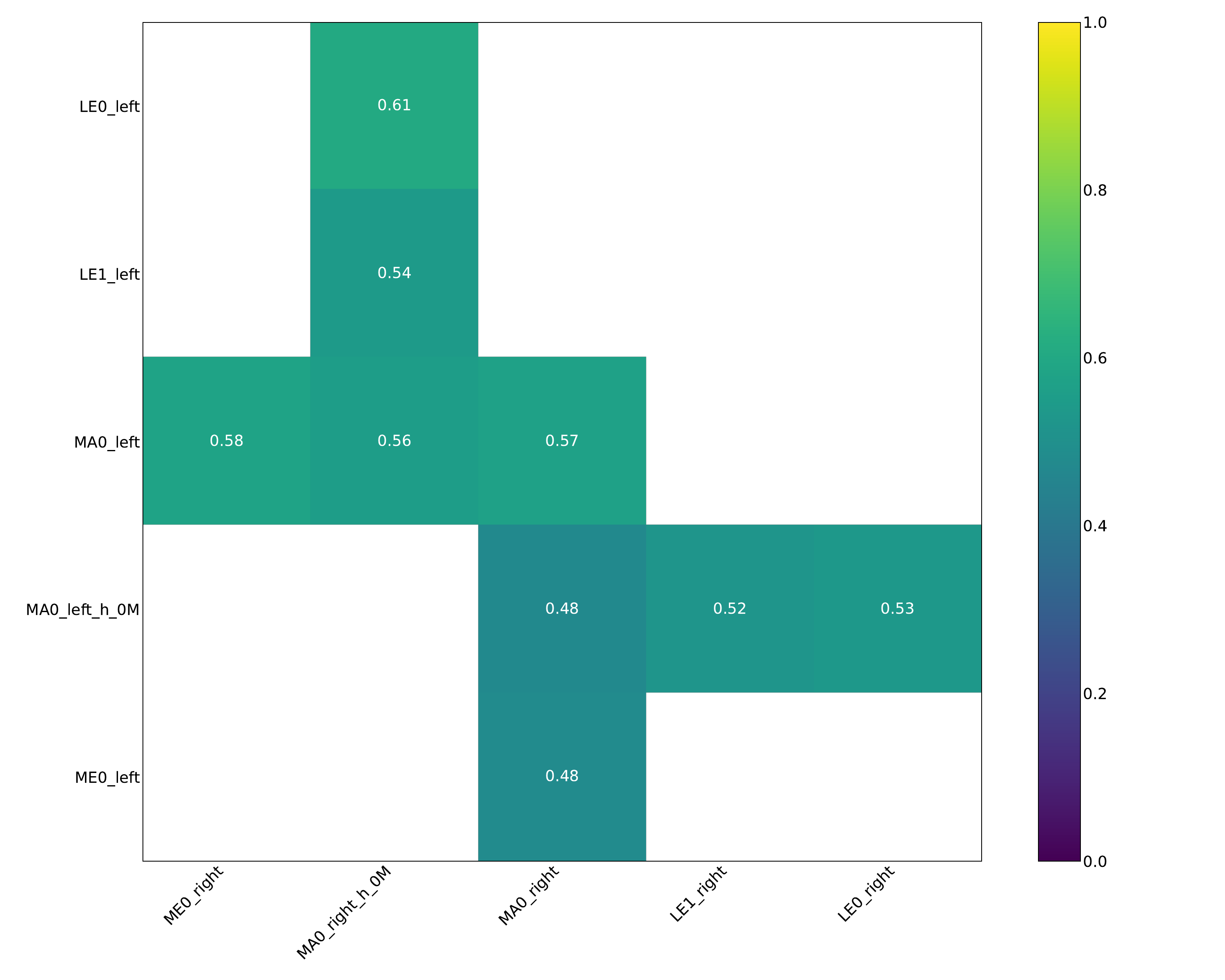}
    \includegraphics[width=0.49\textwidth]{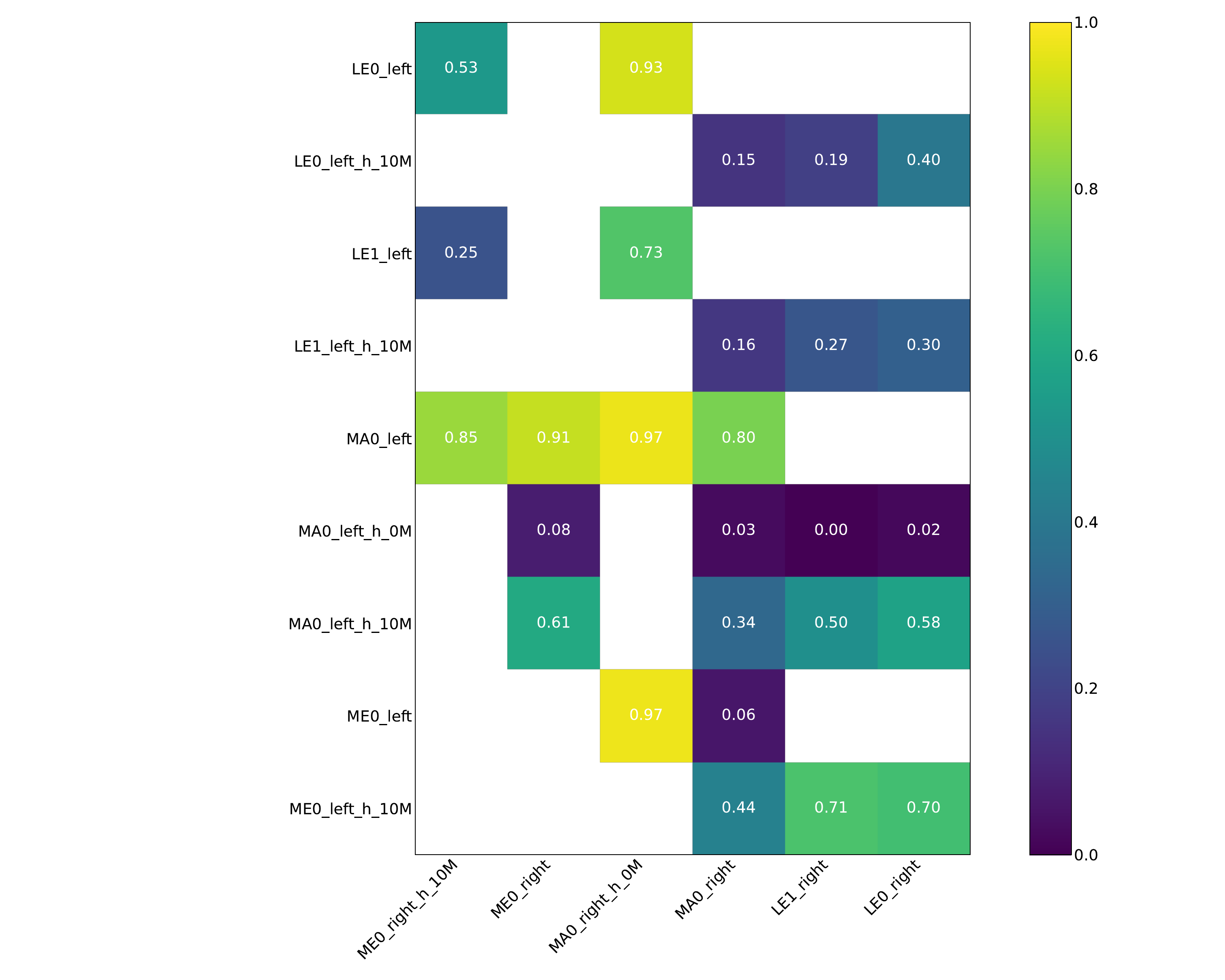}
    \includegraphics[width=0.49\textwidth]{figures/payoff/payoff_lg_3.pdf}
    \includegraphics[width=0.49\textwidth]{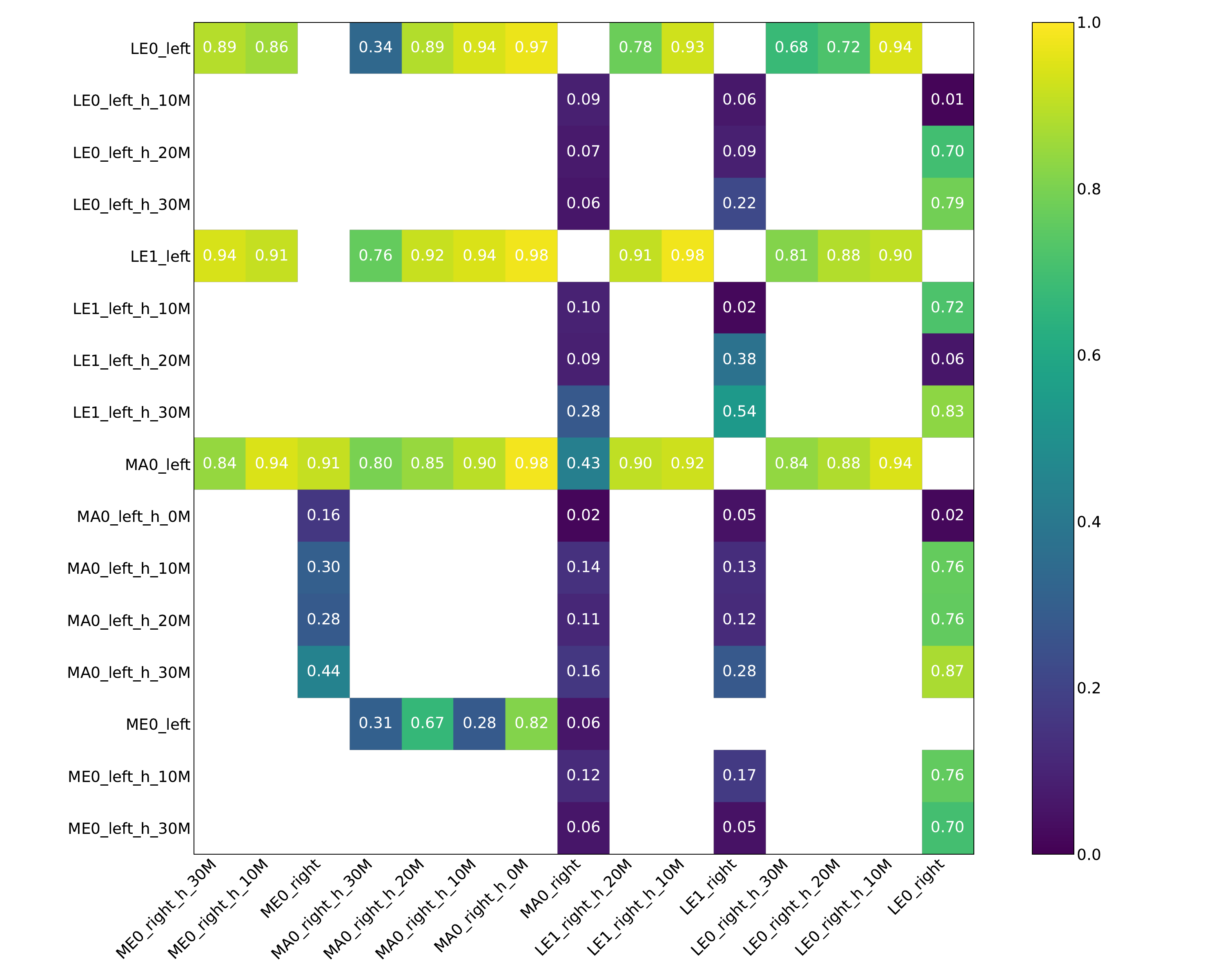}
    \includegraphics[width=0.49\textwidth]{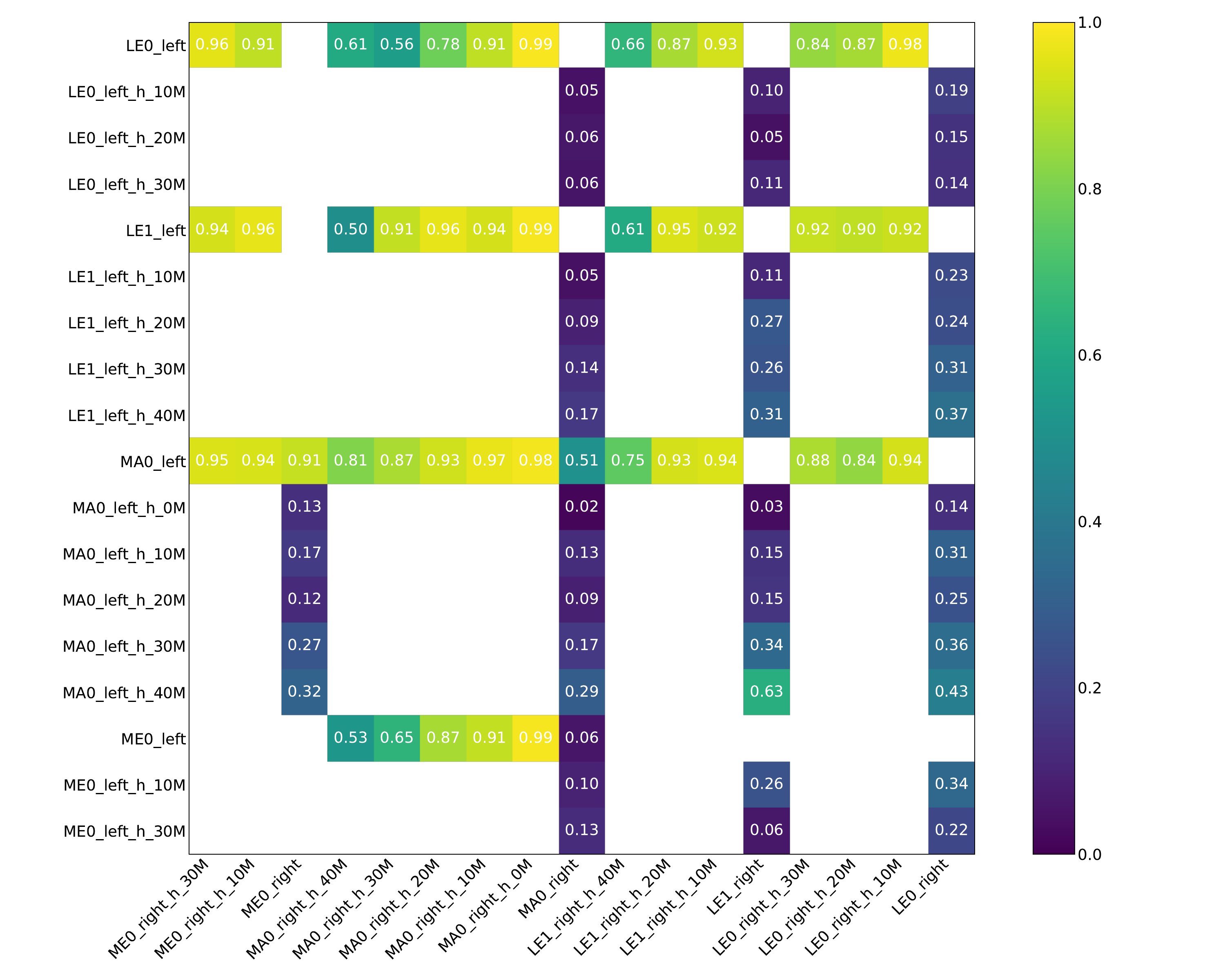}
    \includegraphics[width=0.49\textwidth]{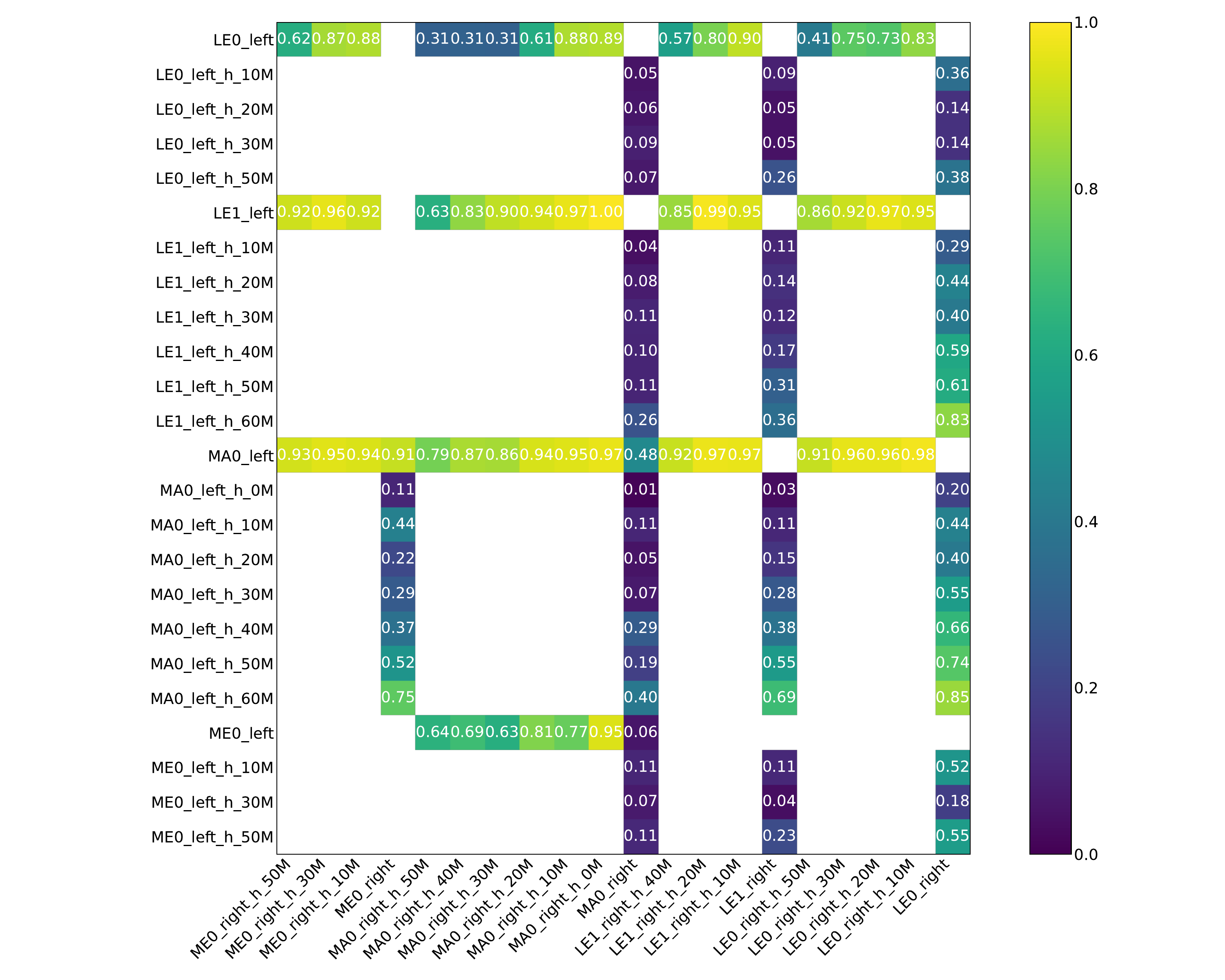}
    \caption{League training details (training order from top left to bottom right): For league training, there is one main agent (MA), two league exploiters (LE0, LE1), and one main exploiter (ME) for each side (left or right). The name of each row indicates the agent information as \texttt{Character\_Side\_Checkpoint}. \texttt{Checkpoint=h\_xM} represents a previous version of agent saved at \texttt{x} million steps. The value indicates the win rate of the left (row) player against the right (column) player.}
    \label{fig:league_payoff}
\end{figure}

Figure~\ref{fig:fsp_payoff}, \ref{fig:psro_payoff}, and \ref{fig:league_payoff} report the payoff matrix of policies within the population for FSP, PSRO, and League, respectively, with the value representing the win rate of the left player against the right player. We trained all our agents on one server with 192 CPUs and 8 A6000 GPUs.

\section{Individual Elo Results} \label{appx:indiv_elo}

\begin{figure}[!h]
    \centering
    \includegraphics[width=\textwidth]{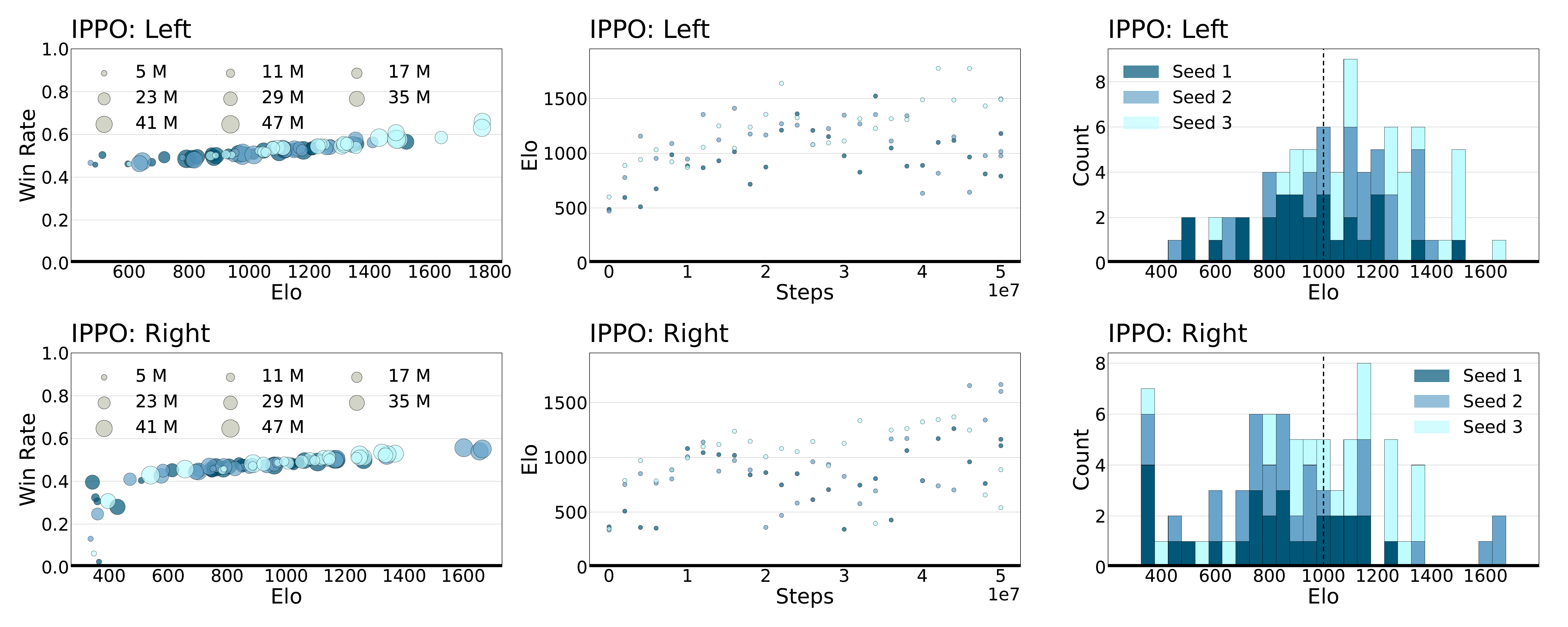}
    \caption{The Elo rating for the population of agents trained with IPPO algorithm. The upper three plots are for left-side player and the bottom three are for the right-side player. The Elo rating is plotted against the winning rate over matched policies (left figures), training steps (middle figures) and the number of policies (right figures).}
    \label{fig:ippo_elo}
\end{figure}

\subsection{IPPO (Figure~\ref{fig:ippo_elo})}

\begin{figure}[!h]
    \centering
    \includegraphics[width=\textwidth]{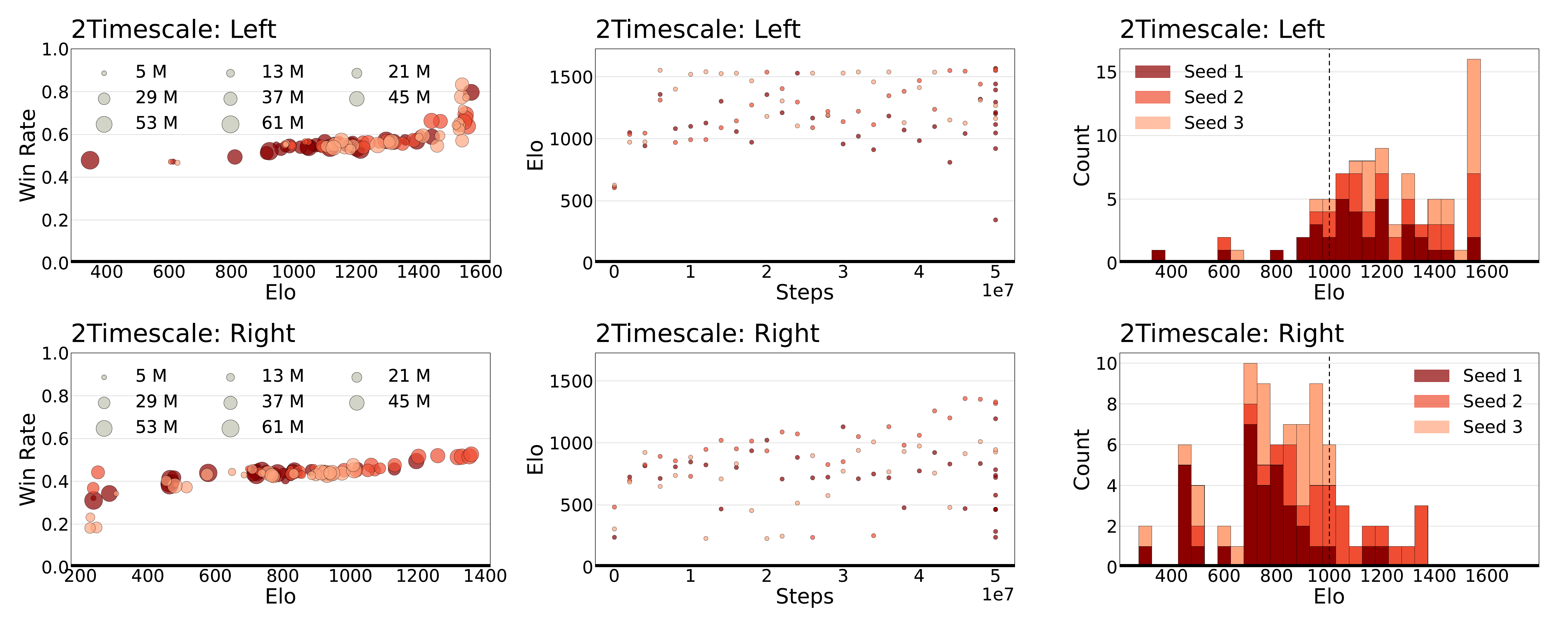}
    \caption{The Elo rating for the population of agents trained with 2Timescale algorithm. The upper three plots are for left-side player and the bottom three are for the right-side player. The Elo rating is plotted against the winning rate over matched policies (left figures), training steps (middle figures) and the number of policies (right figures).}
    \label{fig:2timescale_elo}
\end{figure}

\subsection{2Timescale (Figure~\ref{fig:2timescale_elo})}

\begin{figure}[!h]
    \centering
    \includegraphics[width=\textwidth]{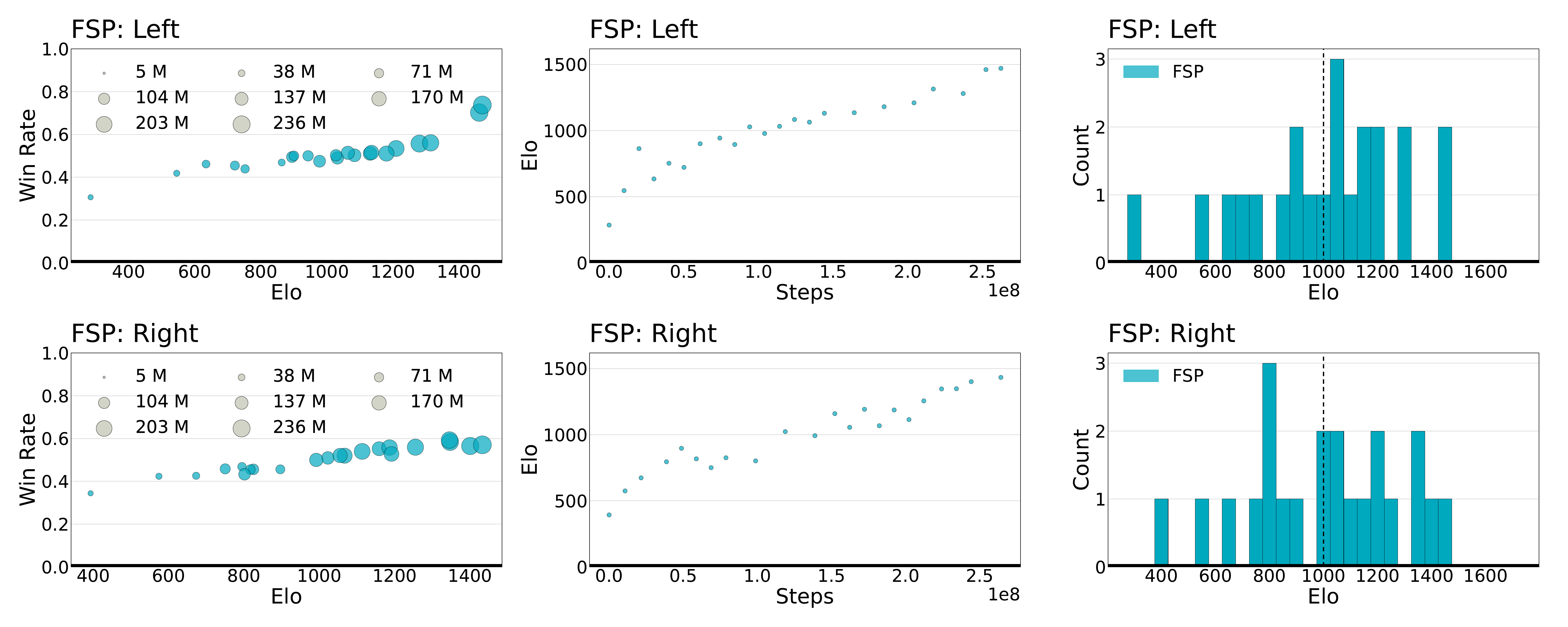}
    \caption{The Elo rating for the population of agents trained with FSP algorithm. The upper three plots are for left-side player and the bottom three are for the right-side player. The Elo rating is plotted against the winning rate over matched policies (left figures), training steps (middle figures) and the number of policies (right figures).}
    \label{fig:fsp_elo}
\end{figure}

\subsection{FSP (Figure~\ref{fig:fsp_elo})}

\begin{figure}[!h]
    \centering
    \includegraphics[width=\textwidth]{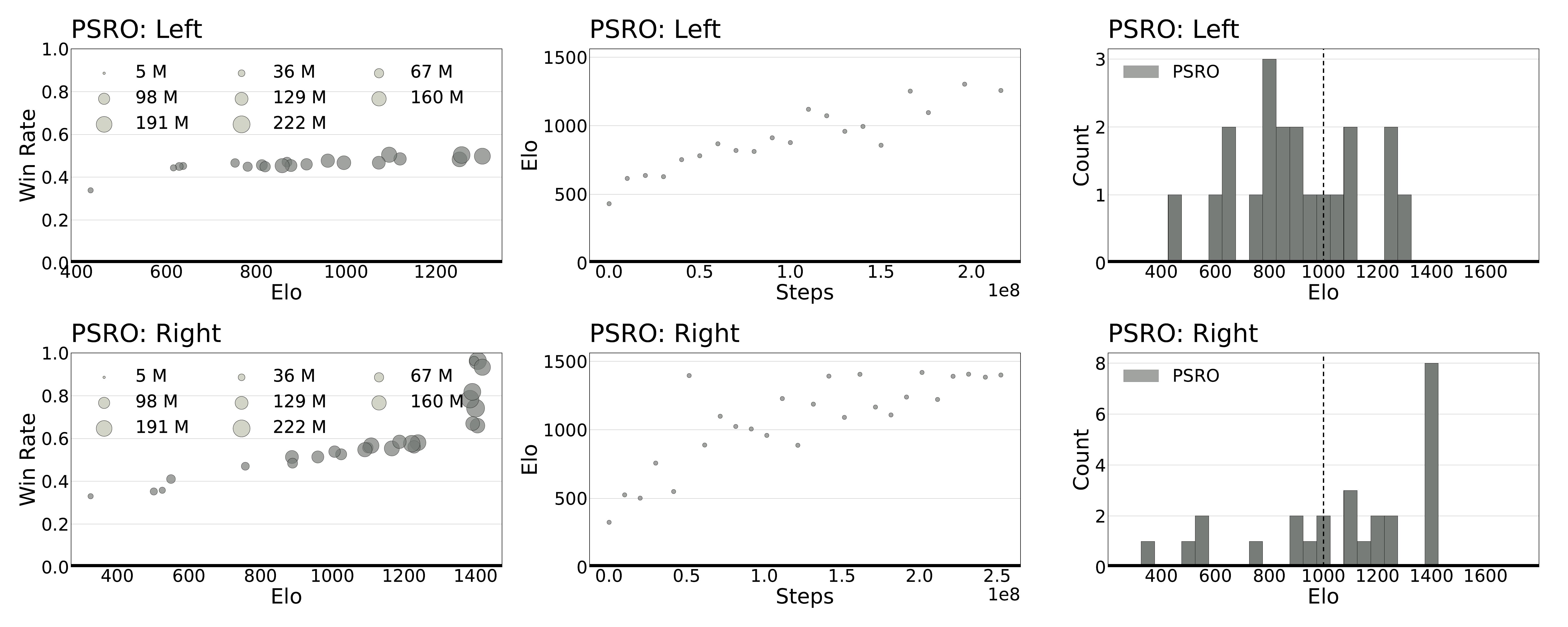}
    \caption{The Elo rating for the population of agents trained with PSRO algorithm. The upper three plots are for left-side player and the bottom three are for the right-side player. The Elo rating is plotted against the winning rate over matched policies (left figures), training steps (middle figures) and the number of policies (right figures).}
    \label{fig:psro_elo}
\end{figure}

\subsection{PSRO (Figure~\ref{fig:psro_elo})}

\begin{figure}[!h]
    \centering
    \includegraphics[width=\textwidth]{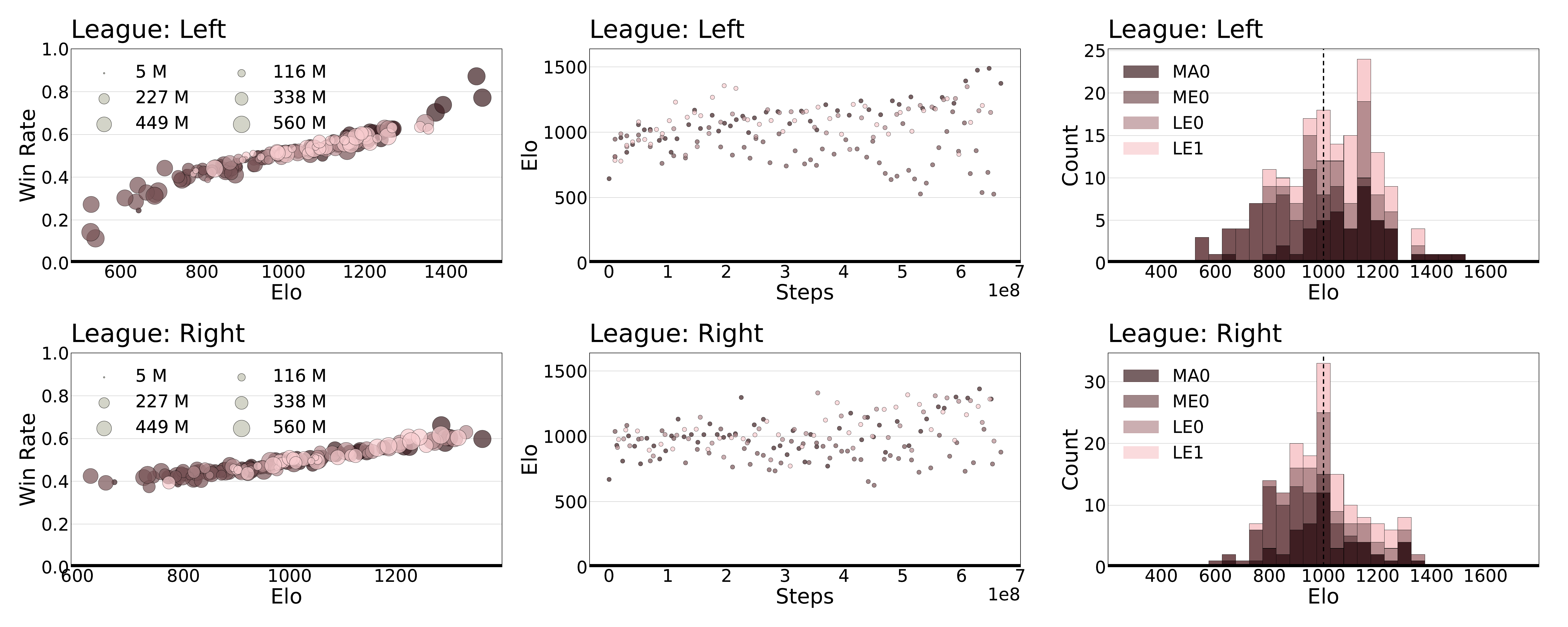}
    \caption{The Elo rating for the population of agents trained with League training. The upper three plots are for left-side player and the bottom three are for the right-side player. The Elo rating is plotted against the winning rate over matched policies (left figures), training steps (middle figures) and the number of policies (right figures).}
    \label{fig:league_elo}
\end{figure}

\subsection{League (Figure~\ref{fig:league_elo})}

\section{Visualization of Human Exploiters} \label{appx:human}
\begin{figure}[!h]
% \vskip 0.2in
\begin{center}
\includegraphics[width=\textwidth]{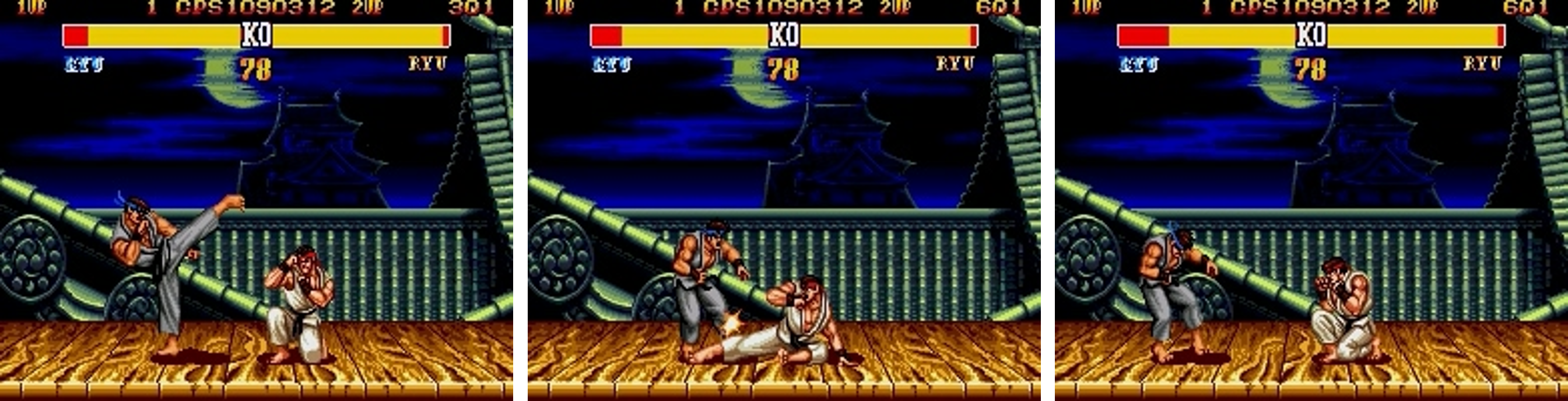}
\vskip -0.1in
\caption{Demonstration of the exploiting strategy of one human player. The human player (Ryu on the right in white) defends when the AI opponent (Ryu on the left in gray) attacks, and inflicts damage with low kicks.}
\label{fig:human}
\end{center}
% \vskip -0.4in
\end{figure}

Figure~\ref{fig:human} visualizes how human players can exploit learned models with a simple strategy. Full videos are provided in the supplementary material. 

\end{document}